\documentstyle[12pt]{article}

\makeatletter

\@addtoreset{equation}{section}
\makeatother
\def\IN{\relax{\rm I\kern-.18em N}}
\def\IR{\relax{\rm I\kern-.18em R}}

\font\cmss=cmss12 \font\cmsss=cmss12 at 7pt
\def\IZ{\relax\ifmmode\mathchoice
{\hbox{\cmss Z\kern-.4em Z}}{\hbox{\cmss Z\kern-.4em Z}}
{\lower.9pt\hbox{\cmsss Z\kern-.4em Z}}
{\lower1.2pt\hbox{\cmsss Z\kern-.4em Z}}\else{\cmss Z\kern-.4em Z}\fi}

\def\inbar{\,\vrule height1.5ex width.4pt depth0pt}
\def\IC{\relax\hbox{$\inbar\kern-.3em{\rm C}$}}

\newcommand{\Od}{{\cal O}}

\newcommand{\Tr}{\mbox{Tr}}

\newcommand{\pabar}{\not{\!{\partial}}}

\newcommand{\Qbar}{\not{\!Q}}
\newcommand{\Abar}{\not{\!\!A}}
\newcommand{\intT}{\int_T d^2x}
\newcommand{\intsp}{\int_{-\infty}^{+\infty} dx} 
\newcommand{\pib}{\int_{periodic} \!\!\!\!\!\!\!\!\!\!d\phi \ }
\newcommand{\pif}{\int_{antiper} \!\!\!\!\!\!\!\!\!\!d\bar\psi d\psi}
\newcommand{\pivec}{\int_{periodic} \!\!\!\!\!\!\!\!\!\!d A_\mu}
\newcommand{\fouri}{\beta^{-1}\sum_{n\in\IZ}
\int_{-\infty}^{+\infty}
 \frac{dp_1}{2\pi}e^{-i p\cdot (x-y)}}
\newcommand{\intk}{\int_{-\infty}^{+\infty} \frac{dk}{2\pi}}
\newcommand{\curr}{j_\mu (x)}
\newcommand{\curreg}{j_\mu^{reg} (x)}
\newcommand{\prodj}{\prod_{j=1}^n}
\newcommand{\prodjN}{\prod_{j=1}^N}
\newcommand{\prodjint}{\prodj\int_T d^2 x_j d^2 y_j}
\newcommand{\sumj}{\sum_{j=1}^n}
\newcommand{\sumn}{\sum_{n=0}^{\infty}}
\newcommand{\sigpm}{\sigma_\pm (x)}
\newcommand{\sigpmr}{\sigma^R_\pm (x)}
\newcommand{\nk}{\left(\begin{array}{c}n\\k\end{array}\right)}
\newcommand{\lims}{\lim_{y\rightarrow x}^s}

\newcommand{\be}{\begin{equation}}
\newcommand{\ee}{\end{equation}}
\newcommand{\ba}{\begin{eqnarray}}
\newcommand{\ea}{\end{eqnarray}}
\newcommand{\dle}[1]{\label{#1}}
\newcommand{\dla}[1]{\label{#1}}
\newcommand{\dr}[1]{\ref{#1}}
\newcommand{\dc}[1]{\cite{#1}}
\newcommand{\dbibitem}[1]{\bibitem{#1}} 
\newcommand{\gsim}{\raise.3ex\hbox{$>$\kern-.75em\lower1ex\hbox{$\sim$}}}
\newcommand{\lsim}{\raise.3ex\hbox{$<$\kern-.75em\lower1ex\hbox{$\sim$}}}

\newcommand{\half}{{\frac{1}{2}}}
\newcommand{\paa}{\partial}

\newcommand{\NP}[1]{{\it Nucl.\ Phys.\ }{\bf #1}}

\newcommand{\PL}[1]{{\em Phys.\ Lett.\ }{\bf #1}}

\newcommand{\AN}[1]{{\em Ann. Phys. } (N.Y.) {\bf #1}}
\newcommand{\CMP}[1]{{\em Comm.\ Math.\ Phys.\ }{\bf #1}}

\newcommand{\PR}[1]{{\em Phys.\ Rev.\ }{\bf #1}}
\newcommand{\PRL}[1]{{\em Phys.\ Rev.\ Lett.\ }{\bf #1}}

\begin{document}

\typeout{--- Title page start ---}

\renewcommand{\thefootnote}{\fnsymbol{footnote}}

\begin{flushright}  
FT/UCM/1-98 \\
DAMTP-1998-145
%\\
%{\tt hep-ph/9810???} \\ 
%\today \\
%\\ (LaTeX-ed on \today )
\end{flushright}
\vskip 12pt

\begin{center} 
{\large\bf Thermal bosonisation in the sine-Gordon and massive
Thirring models}\\
\vskip 1.2cm
{\large
A.G\'{o}mez Nicola$^a$\footnote{E-mail: {\tt Gomez@eucmax.sim.ucm.es}}
and  D.A.Steer$^b$\footnote{E-mail: {\tt D.A.Steer@damtp.cam.ac.uk}}
}\\
\vskip 5pt 
{\it a}) 
Departamento de F\'{\i}sica 
Te\'orica, \\ Universidad Complutense, 28040, Madrid, Spain.
\vskip 3pt
{\it b}) D.A.M.T.P., Silver Street, Cambridge, CB3 9EW, U.K.\\
\end{center}

%\vskip 0.5cm

%\begin{center}
%Tel: +44-1223-338957 \\
%Fax: +44-1223-337918 \\
%\mbox{  }\\
%PACS: 11.10.Wx, 11.10.Kk
%\\
%\mbox{ } \\
%{\bf Keywords} \\
%Finite temperature, chemical potential,2 dimensions
%\end{center}

%See pp.175
\renewcommand{\thefootnote}{\arabic{footnote}}
\setcounter{footnote}{0}
\typeout{--- Main Text Start ---}

\vskip 1cm
\begin{abstract}

We study bosonisation  in the massive Thirring and sine-Gordon models 
at finite temperature $T$ and nonzero fermion chemical potential
$\mu$.  For that purpose we use both canonical operator and path
integral approaches, paying particular attention to the issues of
thermal normal ordering and  renormalisation.  At $T>0$ and $\mu=0$,
the massive Thirring model bosonises to the sine-Gordon model with
the same $T=0$ identification between coupling constants.  We prove
that not only the partition functions of the two models coincide, as
was recently shown, but also that thermal averages of
zero-charge operators can be identified.  In particular, analysis of
the point-split regularised fermion current then leads to the thermal
equivalence between sine-Gordon kinks and Thirring fermions.  At
$\mu \neq 0$, $T>0$ and working in perturbation theory about the
massless Thirring model, we show that the bosonised theory is the
sine-Gordon model plus an additional topological term which accounts
for the existence of net fermion charge excitations (the fermions or
the kinks) in the thermal bath. This result generalises one recently
obtained for the massless case, and it is the two-dimensional
version of the low-energy QCD chiral Lagrangian at finite baryon
density.

\end{abstract}

\vspace{1cm}

\section{Introduction}\dle{sec:intro}

The sine-Gordon (SG) and massive Thirring (MT) models are just two of
many different models which have been widely studied in two space-time
dimensions (2D).  They provide one of the simplest examples of
bosonisation, in which a fermion theory (MT) is equivalent to a
boson one (SG) with certain identifications between their coupling 
constants.  Furthermore the weak limit of one theory is the strong limit
of the other and conversely so that these models also give a simple
example of duality, and the particle spectrum can be mapped from one
model onto the other.  This equivalence was first analysed in
\dc{Coleman} using canonical operator methods and, later on, using
path integral in 
\dc{fugasc82,na85}.  In the MT-SG system, bosonisation takes 
place for the full spectrum.  This should be contrasted, for
instance, to QCD where it happens only at low energies for which
strongly coupled quarks and gluons are confined into hadrons.  There
the effective boson theory is the chiral Lagrangian for the lightest
mesons which are the Nambu-Goldstone bosons of the chiral symmetry
breaking pattern, and to lowest order it is the non-linear sigma
model \cite{dogolope97}. 
 
It is important to note that the particle spectrum of the bosonised
theory may still contain fermion-like excitations.  In the SG model,
such excitations are soliton-like, corresponding classically to
nondissipative solutions of the equations of motion.  One can assign a
conserved charge to those configurations, corresponding to the fermion
charge
\dc{colebook} which is a super-selection rule,  so that the theory can
be classified into different charge sectors.  For instance, a possible 
one-fermion solution is the static kink \dc{Raj}.   The particle
spectrum of the SG theory is then constructed from the mesons
(including kink--anti-kink `breather solutions') and the kinks,
which are in one to one correspondence with the Thirring model 
fermion-antifermion bound states and fermions, respectively. The
situation is entirely analogous to low-energy QCD where the mesons are
the pions, kaons and so on and the solitons are the skyrmions
\dc{sk61}, corresponding physically  to the QCD baryons
\dc{witt83}.  An example in which the bosonised theory does not
contain fermion modes  is the Schwinger model (QED in 2D) which
bosonises to a free massive scalar with mass proportional to the
electric charge \dc{sch62}.  This scalar is a fermion-antifermion
bound state in which the electric charge  is confined. 

The link with QCD discussed above is just one of the motivations to
study these 2D theories  as toy models.  However, one should recall
that some theories in 2D also describe experimentally observed
phenomena. Indeed, the SG equation is behind a device which produces
some of the highest energy microwave signals available, having
technological applications, for example, in  space physics
\dc{JJbook,Roberto1}.  Such Josephson junctions  consist of
two layers of super-conducting materials separated by a very thin
dielectric barrier, and Josephson tunnelling of the  Cooper pairs
across this barrier results in the relative phase angle of the two
superconductors satisfying a SG equation \dc{JJbook}.  As the two
superconductors are taken through the phase transition, kinks are
formed in the junction and these are observed experimentally
\dc{Roberto2}.  The kinks are topological defects, the 2D
analogues of cosmic strings or vortices, line-like defects formed in 4D
when a system goes through a symmetry breaking phase transition (say of
some group $G$ to a subgroup $H$) for which the first homotopy group
$\pi_1(G/H) \neq 0$ \dc{Kib76,ViSh}.  In the early
universe they may provide an explanation for the formation of
structures such as galaxies \dc{ViSh}. An important
problem in this context is to estimate the number density of such
strings formed at the phase transition and also their length
distribution.  Recently progress has been made by using
the analogy between experimentally observable systems such as $^3$He
and $^4$He and the early universe
\dc{GKHe3}.  It would seem, however, that Josephson junctions provide
an even more simple experimentally accessible system with which one
could try to test ideas of defect formation.  Furthermore the analysis
should be simplified in this case by exploiting the duality between
the SG and the MT models \dc{GRS}. For that purpose, 
 one needs to know first 
what is the exact relation between these two models at $T>0$ and $\mu
\neq 0$, and how kink densities and fermion densities are related (see
section \dr{chempot} below).

In this paper we therefore study bosonisation in the SG and MT models
at finite temperature $T$ and non-zero chemical potential $\mu$. 
Recall that regarding bosonisation, temperature and chemical potentials
play a completely different r\^{o}le.  Temperature affects the boundary
conditions of the propagator and hence bosonisation implies a change
from Fermi-Dirac to Bose-Einstein statistics. However, a nonzero
chemical  potential allows the system to have net charged states. As
we have commented before, this is  especially relevant when the
bosonised theory admits fermion states as in the SG model or in chiral
Lagrangians. In fact, we expect the chemical potential to reflect that
property.  Besides, $\mu\neq 0$ spoils the hermiticity of the Dirac
operator and breaks charge conjugation ($C$) invariance. Indeed
breaking the combination of $C$ and Lorentz covariance 
(the thermal bath selects a preferred frame) can give rise to 
topological $\mu$-dependent new terms in the bosonised
action \dc{rewi85,AlvGom95,AlvGom98}.

We now introduce the models.  In Minkowski space with metric $(+,-)$,
the Lagrangian densities for the SG and massive Thirring models are
respectively 
\be
{\cal L}_{SG}[\phi] = \frac{1}{2} \paa_\mu \phi  \paa^\mu \phi +
\frac{\alpha_0}{\lambda^2} \cos \lambda \phi - \gamma_0,
\dle{SG}
\ee
\be   
{\cal{L}}_{MT} [\bar\psi,\psi] =  i\bar{\psi} (\pabar - m_0)\psi +
\frac{1}{2}g^2  j_\mu (x) j^\mu (x)  ,
\dle{Th}
\ee    
where $\curr=\bar\psi (x)
\gamma_{\mu}
\psi (x)$,
$\alpha_0$, $m_0$  and $\gamma_0$ are bare parameters and $\phi$ and
$\psi$ are bosonic and fermionic fields respectively.  The Minkowski 
gamma matrices are
$$
\gamma_0=
\left( \begin{array}{cc} 
   0 & 1 \\   
  1    &   0  
\end{array} \right), \,  \, \, \,    
\gamma_1=\left( \begin{array}{cc}     
    0 & 1 \\      
       -1   &   0  
\end{array} \right) ,   \,  \, \, \,      
\gamma_5 =  \gamma_0 \gamma_1 =
\left( \begin{array}{cc}    
          -1 & 0 \\        
    0    &   1    
\end{array} \right)   
$$  
so that
$$
\{\gamma^\mu,\gamma^\nu\} = 2g^{\mu\nu} 
\; \; \; \; 
\gamma_\mu \gamma_5 =
-\epsilon_{\mu\nu} \gamma^\nu
$$
with $\epsilon_{01}=1$.  

As we have commented before, the SG equations of
motion admit nondissipative solutions for which the classical field 
 $\phi_{cl}(t,\pm\infty)=2\pi
n_{\pm}/\lambda$, with $n_{\pm}\in\IZ$ (with properly normalised
$\gamma_0$). In fact, those classical static solutions with $n_+\neq n_-$
 are the kinks we discussed above. 
%\dnote{Say something more here about different classical
%solutions: kinks, breathers and so on. Also about semiclassical approach,
%although I think that would fit better just below, when we state the way
%we will work, by expanding around the $\alpha=0$ case. The semiclassical  
%would be an alternative approach that we will not attempt here, etc.
% A COUPLE OF REFERENCES WOULD DO} 
It is also worth noticing that the SG Lagrangian in (\dr{SG})
corresponds to a non-linear sigma model in 2D for a single 
Nambu-Goldstone-like field $\phi$. Although there is no spontaneous
symmetry breaking in 2D \dc{col73}, the potential term in (\dr{SG})
breaks explicitly the symmetry $\phi\rightarrow\phi+a$  with $a\in\IR$
(which we will call the chiral symmetry for reasons to become clear
below)  still preserving the symmetry $\phi\rightarrow\phi+2\pi
n/\lambda$ with $n\in\IZ$. These two symmetries are,  respectively, the
counterparts of the  chiral and isospin symmetries for QCD, $\alpha_0$ and
$\lambda$ playing the r\^ole of the pion mass squared and the inverse
of the pion decay constant respectively.  On the other hand, the chiral
symmetry transformations in terms  of the Thirring fermion are
$\psi\rightarrow\exp(ia\lambda\gamma_5)\psi$.  The massless Thirring model is
chirally invariant, the fermion  mass term breaking that symmetry
in the same way as the $\alpha$ term does in the SG Lagrangian.

We now recall the results at $T=\mu=0$.  In \dc{Coleman}, Coleman, 
using a canonical field operator approach, considered chiral invariant
combinations of zero-charge operators\footnote{These operators have
the same number of fermions and anti-fermions in the MT, and
correspond to operators which are invariant under $\phi\rightarrow\phi
+2\pi n/\lambda$, with $n\in\IZ$, in the SG model.} 
and showed that the SG and the
massive Thirring models are equivalent in the following sense
\ba
m \bar\psi(x)\psi(x) & = & \frac{\alpha}{\lambda^2}
\cos\lambda\phi(x),
\dla{equivoper1}
\\
 \curr & = &\frac{\lambda}{2\pi}
\epsilon_{\mu\nu} \partial_\nu\phi(x),
\dla{equivoper2}
\ea
provided  
the constants of the two models are related through
\ba
\frac{\lambda^2}{4\pi}&=&\frac{1}{1+g^2/\pi},
\dla{equivconst1}
\\
\frac{\alpha}{\lambda^2}&=&\rho m ,
\dla{equivconst2}
\ea
where $\rho$ is the renormalisation scale and all the coupling
constants and fields in (\dr{equivoper1})-(\dr{equivoper2}) are 
renormalised (see sections \dr{subs:sgth} and \dr{ren}). 
In fact the renormalisation convention chosen here is not exactly 
the same as that of \dc{Coleman} (see below) and
it also differs from \dc{zj} by a $\pi$ factor which may be absorbed
in $\rho\rightarrow\rho/\pi$ in the above equation.  All the physical 
results must be  independent of the scale.  Throughout this paper, we 
will refer to (\dr{equivoper1}) and (\dr{equivoper2}) as the mass and current
equivalences respectively.  It is very important to observe that
whilst it may be inferred from
\dc{Coleman} that the above equivalences are strong, this is {\em
not} the case---they are
weak identities, {\em only} holding between the vacua of the two
theories.  We also recall that (\dr{equivoper2}) allows one to
establish that the net fermion charge (fermions minus anti-fermions)
in the massive Thirring model corresponds to the net kink number
(kinks minus anti-kinks) in the SG model\footnote{Note that here we are 
calling quantum configurations with any nonzero net charge and 
space-time dependence kinks, although strictly speaking the kinks are just
time-independent classical configurations with net charge equal to one.}  
\dc{Coleman,colebook}.
%\dnote{I WOULD CUT THIS SENTENCE --- IT REPEATS
%TOO MUCH:  As we
%have commented before, the fundamental fermion excitations of the
%massive Thirring model are the kinks of the SG model, whereas the
%fundamental meson excitations of the SG model are fermion-antifermion
%bound states in the Thirring model.}
  Equations
(\dr{equivconst1})-(\dr{equivconst2}) were also obtained using path
integral techniques in \dc{fugasc82,na85}.
In both \dc{Coleman} and \dc{fugasc82,na85}, the approach followed was
to expand formally around the $\alpha_0=0$ in the SG model and the
$m_0 =0$ in the Thirring model---we will do the same here.  An
alternative approach would be, for instance, to work 
within a semiclassical approximation \cite{Coleman,goja75}, 
which we will not  attempt in this 
work. 

A question which may be asked is whether the equivalences
(\dr{equivoper1})-(\dr{equivoper2}) also hold in a heat bath where the
vacuum expectation values are replaced by thermal expectation values. 
 For bosons,
\be
\ll \bullet \gg \; = \frac{1}{Z(T)}\Tr\left(\bullet 
e^{-\beta \hat H}\right)=
\frac{N_\beta}{Z(T)}\pib \bullet \exp (-S[\phi]),
\dle{thav}
\ee
where $N_\beta$ is an infinite $T$-dependent constant arising in the path
integral description \dc{ber74}, $\beta=1/T$, the path integral 
boson fields are periodic in Euclidean time
$\phi(\bar{t}+\beta)=\phi(\bar{t})$ with $\bar{t} = i t$ and
$$
Z(T)=\Tr \left( e^{-\beta \hat H} \right)
=  N_\beta  \pib  \exp (-S[\phi])
$$
is the partition function. Thermal averages and partition functions 
for fermions  are defined analogously, but integrating over anti-periodic
field configurations instead. It is important to remember that for
$\mu=0$, one  works in the canonical ensemble, fixing the net fermion
charge to zero.  Consistently, in that case, the above trace is taken in
the SG model over all possible states with zero net kink number or,
equivalently, over field configurations that vanish at spatial
infinity (configurations such as a kink-antikink pair are
permitted).  On the other hand, for $\mu\neq 0$, we are working in the
grand-canonical ensemble and the net fermion number is not fixed.  Hence
the net averaged fermion density is nonzero and we are summing over all
possible states in all the super-selection sectors of the theory, i.e,
over all possible net kink numbers in the SG model. In other words, the
complete set of states over which the trace is taken can be chosen to
be eigenstates of the Hamiltonian for $\mu=0$ and eigenstates of both the
Hamiltonian and the number operator for $\mu\neq 0$.  
%\dnote{CUT:  The answer to the above question does not seem to be
%immediately obvious, since thermal field theory is a non-trivial
%combination of QFT with Statistical Mechanics, where concepts such as
%asymptotic states are not well defined.  Therefore, it is not entirely
%clear how on-shell information from the zero-temperature $S$-matrix
%elements of the two models can be applied to the finite temperature
%equilibrium case where the $S$-matrix is in fact unity.}
%\dnote{If we decide to keep this last sentence, we should give a
%reference. However, it is not clear to me that we should say this,
%because there are works, such as Dashen et al that do claim that the
%partition function can be obtained from the $S$-matrix elements} 
%
In \dc{Belg} path integral methods were used to show that the SG and
MT model partition functions are equivalent at $T>0$ but
$\mu=0$. On the other hand, it was shown in \dc{AlvGom98} that 
for $m_0=0$ and $\mu \neq 0$, the Thirring model partition function is not the
free boson one ($\alpha_0=0$), but it acquires an extra $\mu$-dependent
term accounting for the nonzero fermion net number in the thermal
bath. 

In this work we analyse several aspects of the thermal
bosonisation  for this system which we believe have not been studied
before.  In the first part of the paper (section \dr{sec:op}) we
extend the work of Coleman
\dc{Coleman} to $T>0$ and $\mu=0$, working  with a canonical operator
approach.   Particular attention is paid to the question of
renormalisation and  the  definition of  normal ordering in a heat
bath.  In the process, the results of \dc{Belg} are reproduced though
the approach is entirely different.  The rest of the paper uses path
integral methods, which are introduced in section \dr{sec:pi}.  There 
we again prove that the partition functions of the
two models are the same, but we 
pay special attention to renormalisation and to the connection with our
results in the canonical formalism.  The main objective of section
\dr{sec:piopequiv} is to prove that thermal averages  of
correlators of the zero-charge operators 
(\dr{equivoper1})-(\dr{equivoper2}) evaluated at different space-time
points coincide.  Such correlators cannot be inferred in general from
the partition function which contains information about global
thermodynamic observables such as the pressure or the condensates, but
not about correlators, which physically yield for example thermal
correlation lengths. In our case,  the thermal version of
(\dr{equivoper2}) will constitute the thermal Thirring fermion-SG
kink equivalence. This equivalence will appear again in section
\dr{chempot}, where we analyse  the case $T>0$ and $\mu\neq 0$ for the
partition function, thus extending \dc{AlvGom98} to the massive case
and \dc{Belg} to $\mu\neq 0$.  The massless and $\mu=0$ case was 
 analysed in \dc{rualv87}. A summary of the results presented in
this paper may be found in \dc{TFT98}.

\section{Bosonisation in the canonical operator approach at $T>0$ and
$\mu=0$}\dle{sec:op}

Here we work in Minkowski space with a canonical operator approach
and extend the results of \dc{Coleman} to $T>0$ through the
use of thermal normal ordering introduced in \dc{ES} and defined
below.  Calculations are carried out in perturbation theory, expanding
about $\alpha_0 = 0$ in the SG model and about $m_0=0$ in the Thirring
model.  We first summarise useful results for
2D free partition functions and propagators at finite temperature and
zero chemical potential.

\subsection{Results for free fields in 2D at
$T>0$ and $\mu=0$.}\dle{subs:res}

\subsubsection{Notation and free partition functions}\dle{ssubs:pf}

For simplicity we take the spatial
dimension to be finite $0 \leq x \leq L$ with corresponding
discrete momenta 
$k := k_n = 2 \pi n/L$.  The infinite length limit is reached by
substituting $\sum_{ k} \longrightarrow  L \intk $.  As opposed to the
fields in the path integral (sections \dr{sec:pi}-\dr{chempot}), field
operators  are entirely independent of temperature $T = \beta^{-1}$ and
hence do not satisfy any periodicity conditions in imaginary
time\footnote{Temperature dependence, and thus questions of periodicity,
only appear when considering thermal expectation values of operators.}. 
As usual, all the thermodynamic observables are obtained from the
partition function and its derivatives by dividing by $\beta L$.

Consider a free real scalar field $\hat{\phi}$ of mass $\tilde{\mu}$ and a
free massless fermion field $\hat{\psi} $ with corresponding Hamiltonians
\be
\hat{H}_0^{\tilde{\mu}} = \int_{0}^{L}d x \left[ 
\frac{\hat{\pi}^2}{2} + \frac{1}{2} \left( \frac{\paa \hat{\phi}}{\paa
x}
\right)^2 + \frac{\tilde{\mu}^2 \hat{\phi}^2}{2} \right];
\; \; \; \; \; \; \; 
\hat{H}_0^{F} = - i \int_{0}^{L}d x  \left[ \overline{\hat{\psi}} \gamma^1
\paa_1
\hat{\psi} \right]
\dle{Hofermdef}
\ee
where $\hat{\pi} = \paa_0 \hat{\phi}$, $\overline{\hat{\psi}} =
\hat{\psi}^{\dagger} \gamma_0$ and the subscript $0$ denotes the free case.
As usual, the free field operators (or more generally operators in the
interaction picture) are written in terms of annihilation and creation
operators as
\ba
\hat{\phi}(t,x) &=& \sum_k \; (2\omega_{k,\tilde{\mu}} L)^{-1/2} 
\left[ \hat{a}_{k,\tilde{\mu}} e^{-ik\cdot x} + 
\hat{a}^{\dagger}_{k,\tilde{\mu}} e^{ik\cdot x}
\right],
\dla{phidef}
\\
\hat{\psi}(t,x) &=&   \sum_k \; L^{-1/2}
\left[ \hat{c}_{k} e^{-ik\cdot x} + \hat{d}^{\dagger}_{k} e^{ik\cdot x}
\right] u(k),
\nonumber
\ea
where $\omega^2_{k,\tilde{\mu}}=k^2 + \tilde{\mu}^2$ and the 
two-vector $u(k)$ is
$$
u(k) =  \left( \begin{array}{c} 
   \theta(-k) \\   
  \theta(k) 
\end{array} \right).
$$
Also $k\cdot x$ is shorthand for $k_0 t
- k x$ with $k_0=\omega_{k,\tilde\mu}$, and 
$$
\left[ \hat{a}_{k,\tilde{\mu}},\hat{a}^{\dagger}_{k',\tilde{\mu}} \right] = 
\left\{ \hat{c}_{k},\hat{c}^{\dagger}_{k'} \right\} = 
\left\{ \hat{d}_{k},\hat{d}^{\dagger}_{k'} \right\} =\delta_{k,k'} 
$$
with all other commutators and anti-commutators vanishing.  Note that
for the scalar field, our notation differs from the conventional one
in that the annihilation and creation operators carry an extra
label which is the mass $\tilde{\mu}$ appearing in the free Hamiltonian.
In terms of annihilation and creation operators, the boson Hamiltonian
is as usual
\be
\hat{H}_0^{\tilde{\mu}} = \sum_{k} \frac{\omega_{k,\tilde{\mu}}}{2} 
\left[\hat{a}_{k,\tilde{\mu}} \hat{a}^{\dagger}_{k,\tilde{\mu}}
+ \hat{a}^{\dagger}_{k,\tilde{\mu}} \hat{a}_{k,\tilde{\mu}} \right] 
= \; 
:\hat{H}_0^{\tilde{\mu}} :_{\tilde{\mu}} + \sum_{k}
\frac{\omega_{k,\tilde{\mu}}}{2} ,
\dle{H0}
\ee
where the standard normal ordering operation, $: \bullet:_{\tilde{\mu}}$,
which places the annihilation operators
$\hat{a}_{k,\tilde{\mu}}$ to the right of
creation operators $\hat{a}^{\dagger}_{k,\tilde{\mu}}$, has been 
used to separate out the infinite vacuum energy.  Note again the mass
label on the normal ordering operation---it will be important below where
objects such as $:\hat{H}_0^{\tilde{\mu}}:_{\rho}$ with $\rho \neq
\tilde{\mu}$ are considered.  (By this we mean that the operators
$\hat{\phi}$ must be expressed in terms of a mass $\rho$ in (\dr{phidef}),
and then the creation and annihilation operators
$\hat{a}_{k,\rho}$ are ordered.  This is equivalent to considering a mass
shift in the Hamiltonian:  $\hat{H}_0^{\tilde{\mu}} =
\hat{H}_0^{\rho} + (\tilde{\mu}^2 - \rho^2) \hat{\phi}^2/2$.)   From
(\dr{H0}), the free boson partition function is, in the limit $L
\rightarrow \infty$,
\be
\log Z_0^{\tilde{\mu},B}(T) =
\log \left[ \Tr \left\{ e^{-\beta \hat{H}_0^{\tilde{\mu}}} \right\} \right]
= -L\intk\left[\frac{\beta\omega_{k,\tilde{\mu}}}{2}+
\log\left(1-e^{-\beta\omega_{k,\tilde{\mu}}}\right)\right].
\dle{partbos}
\ee
The first term in the far r.h.s  above is divergent and corresponds to the 
vacuum energy.
In the canonical approach it is renormalised by normal
ordering.  In path-integral methods there is no operator normal
ordering and infinite vacuum terms are not removed, although they are 
 irrelevant for the thermodynamics (they are $T$-independent) 
and can be ignored \dc{ber74}. 
  For 
the massless fermion hamiltonian one has
$$
\hat{H}_0^{F} =  \sum_{k} |k|
\left( \hat{c}^{\dagger}_{k} \hat{c}_{k} - 
 \hat{d}_{k} \hat{d}^{\dagger}_{k} \right) 
 =  \sum_{k} |k|
\left( \hat{c}^{\dagger}_{k} \hat{c}_{k} +
\hat{d}^{\dagger}_{k} \hat{d}_{k} \right) -  \sum_{k} |k|
= \; :\hat{H}_0: -  \sum_{k} |k| ,
$$
whence 
\be
\log Z_0^{F} =2 L \intk \left[\frac{\beta \vert k\vert}{2}+
\log\left(1+e^{-\beta \vert k\vert}\right)\right].
\dle{partferm}
\ee
Note that since
$$
\intk \log\left(1-e^{-\beta \vert k\vert}\right)=
-2\intk \log\left(1+e^{-\beta \vert k\vert}\right)=-\frac{\pi T}{6},
$$
it follows that, ignoring vacuum
terms, 
\be
Z_0^{F}(T) = Z_0^{0,B}(T) = \exp {\left[\frac{\pi L
T}{6}\right]} .
\dle{pffree}
\ee
This result provides one of the simplest examples of
bosonisation, and we will make use of it later on since it forms the
basis of the equivalence of the SG and MT partition functions at $T>0$.

\subsubsection{Thermal normal ordering and thermal
propagators}\dle{ssubs:tno}

Central to our discussion below will be the use of `thermal
normal ordering' \dc{ES}.  The reason one is lead to consider a
different normal ordering at $T>0$ comes from the fact that whilst at
zero temperature, dealing with normal ordered products is easy as
by construction their vacuum expectation value vanishes, for
example,
\be
\langle 0,\tilde{\mu}| : \hat{a}_{k,\tilde{\mu}} 
\hat{a}_{k,\tilde{\mu}}^{\dagger} :_{\tilde{\mu}} |0,\tilde{\mu} 
\rangle = 0,
\dle{T0NO}
\ee
this is no longer the case at $T>0$ where
\be
\ll : \hat{a}_{k,\tilde{\mu}} \hat{a}_{k,\tilde{\mu}}^{\dagger} 
 :_{\tilde{\mu}} \gg_0 \; = 
\frac{
\sum_{n=0}^{\infty} \langle n,\tilde{\mu}|  e^{-  \beta
:\hat{H}_{0}^{\tilde{\mu}}:_{\tilde{\mu}}  } 
\hat{a}_{k,\tilde{\mu}}^{\dagger}\hat{a}_{k,\tilde{\mu}}
 | n,\tilde{\mu} \rangle
}
{\sum_{n=0}^{\infty}\langle n,\tilde{\mu}|
e^{-\beta :\hat{H}_{0}^{\tilde{\mu}}:_{\tilde{\mu}} } 
 | n,\tilde{\mu} \rangle} =
\frac{1}{e^{\beta
\omega_{k,\tilde{\mu}}} -1 }= N_{k,\tilde{\mu}} \neq 0. 
\dle{defES}
\ee
Here $N_{k,\tilde{\mu}}
%= (e^{\beta
%\omega_{k,\tilde{\mu}}}- 1)^{-1}
$ is the Bose-Einstein distribution. 
As was discussed in \dc{ES,EKS}, this problem can be avoided by
introducing a more general and indeed more convenient definition of
normal ordering than the one given above.  It is obtained by splitting
the field operators into two {\em arbitrary} parts $\hat{\Psi} =
\hat{\Psi}^+ +
\hat{\Psi}^-$ ($+$/$-$ do {\em not} refer to positive/negative energy
waves) with normal ordering
$N^{ES}$ defined to place the `positive' parts to the right of the
`negative' parts so that, for example
\be
N^{ES} \left[ \hat{\Psi}_1 \hat{\Psi}_2 \right] =
\hat{\Psi}^{+}_1 \hat{\Psi}^{+}_2 +
\hat{\Psi}^{-}_1 \hat{\Psi}^{+}_2 + \sigma \hat{\Psi}^{-}_2  
\hat{\Psi}^{+}_1+
\hat{\Psi}^{-}_1 \hat{\Psi}^{-}_2
\dle{N2}
\ee
where $\hat{\Psi}_i = \hat{\Psi}(x_i)$, $\sigma = 1$ for bosons 
and $\sigma=-1$ for fermions.
If the contraction is defined in the usual way in terms of the
two-point time ordered product and normal ordered product (\dr{N2}),
then the operator form of Wick's theorem holds in its usual form 
provided the contraction is a $c$-number \dc{EKS}.  In thermal field
theory, one can then {\em choose} a split which
satisfies
\be
\ll N^{ES} [\bullet] \gg_0 \; = 0
\dle{NOcond}
\ee
for all non-constant operators $\bullet$ in the interaction picture, thus
mimicking the $T=0$ case (\dr{T0NO}).  For boson fields and all contours
$C$ used in thermal field theory \dc{lebe96}, this split is given by
\dc{ES} 
\ba
\hat{\phi}^+ &=& \sum_k \;  (2\omega_{k,\tilde{\mu}} L)^{-1/2} 
\left[ (1-f_{k,\tilde{\mu}}) \hat{a}_{k,\tilde{\mu}} e^{-ik\cdot x} + 
g_{k,\tilde{\mu}} 
\hat{a}^{\dagger}_{k,\tilde{\mu}} e^{ik\cdot x}
\right]
\dla{phipl}
\\
\hat{\phi}^- &=& \sum_k \; (2\omega_{k,\tilde{\mu}} L)^{-1/2} 
\left[ f_{k,\tilde{\mu}} \hat{a}_{k,\tilde{\mu}} e^{-ik\cdot x} + 
(1-g_{k,\tilde{\mu}}) 
\hat{a}^{\dagger}_{k,\tilde{\mu}} e^{ik\cdot x}
\right]
\dla{phiminus}
\ea
where
\be
(1-f_{k,\tilde{\mu}})(1-g_{k,\tilde{\mu}})=1+N_{k,\tilde{\mu}},
  \; \; \; \; \; 
f_{k,\tilde{\mu}} g_{k,\tilde{\mu}}=-N_{k,\tilde{\mu}} 
\dle{fdef}.
\ee
Similar expressions hold for fermion field operators 
\dc{ES} and also for other operators such as the momenta operator
$\hat{\pi} $.  Note that
$N^{ES}[\bullet]$, the operation we call thermal normal ordering
(TNO) when (\dr{NOcond}) is satisfied, 
reduces to $:\bullet:$ at $T=0$.

We will need the free thermal boson propagator
$$
\Delta_{T}(x) = \; \ll T_c \left[ \hat{\phi}(x) \hat{\phi}(0)
\right] \gg_0 
= \theta_c(t) \left[
\hat{\phi}^+(x),\hat{\phi}^-(0)\right] + \theta_c(-t) \left[
\hat{\phi}^+(0),\hat{\phi}^-(x)\right]
$$
where $T_c$ means contour ordering.
% along a contour $C$ in complex time that 
%runs from some $t_i$ to $t_i-i\beta$ .  
In the imaginary time
formalism (with the contour starting at $t_i = 0$ and ending at $-i
\beta$), use  of (\dr{phipl})-(\dr{fdef})
gives
\be
\Delta_T(x)=-\frac{1}{4\pi}\log \tilde{\mu}^2\beta^2 Q^2(x)+ K +
\Od(\tilde{\mu}\beta).
\dle{bosprop}
\ee
where we have expanded in powers of $\beta \tilde{\mu} \ll 1$ and $K$
is a constant, which will not play any r\^{o}le in our analysis (see below).
 The $Q$ variable is given by $Q^2=Q_0^2+Q_1^2$ where
\ba
Q_0(x,\bar{t})&=&-\cosh (\frac{\pi x}{\beta}) \sin (\frac{\pi
\bar{t}}{\beta})
\\ Q_1(x,\bar{t})&=&-\sinh (\frac{\pi x}{\beta}) \cos
(\frac{\pi \bar{t}}{\beta})
\dla{qs}
\ea
so that $Q(x)$ is a Lorentz scalar.
Observe that the boson propagator (\dr{bosprop})
is both ultra-violet (UV) divergent ($x \rightarrow 0$) as well as
infra-red (IR) divergent ($\tilde{\mu}\rightarrow 0^+$).  It also has the
same form as the zero temperature boson propagator, as for both
$T\rightarrow 0^+$ and $x \rightarrow 0^+$, $Q_\alpha\rightarrow \pi
Tx_\alpha$, so that the $T=0$ propagator is recovered from (\dr{bosprop}) 
by replacing $Q_\alpha\rightarrow T x_\alpha$
%\dnote{This must be somehow a
% consequence of the conformal invariance of the free theory, 
% should we say something or cite 
% \cite{zj} for instance?}. 
 For the remainder of this
section we will label the propagator also by its mass $\tilde{\mu}$ so
that in (\dr{bosprop}), $\Delta_T(x) \rightarrow \Delta_T(x;\tilde{\mu})$.

We note here other useful properties of the $Q$ variables that will 
be needed later on. The first one concerns their short and long 
 distance behaviour, which, from (\dr{qs}) is
\ba
Q^2(x,\bar{t})&\stackrel{\vert x\vert\rightarrow\infty}
{\longrightarrow}&\frac{1}{4}
e^{ 2\pi \vert x\vert /\beta} \qquad \forall \bar{t}\nonumber\\
&\stackrel{(x,\bar{t})\rightarrow (0,0)}{\longrightarrow}&(\pi T)^2
(x^2+\bar{t}^2)
\dla{asbeq}
\ea
 and  the second one is
\ba
\epsilon^\mu\frac{\partial}{\partial x^\mu}\left[
\frac{1}{\epsilon_0+i\epsilon_1}\log Z^+_Q (x)-
 \frac{1}{\epsilon_0-i\epsilon_1}\log Z^-_Q (x)\right]&=&
-i\frac{\partial}{\partial x^1}\log Q^2(x) 
\nonumber\\
\epsilon^\mu\frac{\partial}{\partial x^\mu}
\left[\frac{1}{\epsilon_0+i\epsilon_1}\log Z^+_Q (x)+
 \frac{1}{\epsilon_0-i\epsilon_1}\log Z^-_Q (x)\right]&=&
\frac{\partial}{\partial x^0}\log Q^2(x) 
\dla{propQ}
\ea
with
\be
Z_Q^\pm (x)=Q_0 (x)\pm i Q_1 (x).
\dle{zq}
\ee

%\dnote{Dani thinks we might cut the whole of this paragraph.  In
%particular if we decide to keep it then we have to define normal
%ordering with the subscript $\tilde{\mu}$ which as yet is only done in
%the next section after the SG hamiltonian.}We
%conclude this section by commenting that while TNO is only really
%useful when calculating thermal expectation values of fields (see 
%below),
%it is interesting to note that the thermally normal ordered Hamiltonian
%(\dr{Hofermdef}) is given by
%$$
%N^{ES}_{\tilde{\mu}} [\hat{H}_0^{\tilde{\mu}} ] =  \sum_{k}  
%\omega_{k,\tilde{\mu}}
%\left[ \hat{a}_{k,\tilde{\mu}}^{\dagger} \hat{a}_{k,\tilde{\mu}} 
%- N_{k,\tilde{\mu}} \right]
%$$
%so that 
%$$
%\hat{H}_0^{\tilde{\mu}} =   
%N^{ES}_{\tilde{\mu}} [\hat{H}_0^{\tilde{\mu}} ]  +  \sum_{k}  
%\frac{\omega_{k,\tilde{\mu}}}{2} \left( 1
%+ 2 N_{k,\tilde{\mu}} \right) .
%$$
%This is, after all, the expected result as $\sum_{k}  
%\frac{\omega_{k,\tilde{\mu}}}{2} \left( 1
%+ 2 N_{k,\tilde{\mu}} \right) = \; \ll \hat{H}_0^{\tilde{\mu}} 
%\gg_0$ is just the
%average thermal energy, the natural generalisation of the zero
%temperature infinite vacuum energy $\langle 0,\tilde{\mu}| 
%\hat{H}_0^{\tilde{\mu}}|
%0,\tilde{\mu}
%\rangle$.  Here we see that thermal normal ordering has separated
%out the usual infinite vacuum term as well as an extra finite part
%which is the finite contribution to the thermal energy.  In the
%case of zero mass one also has
%$\Tr
%\left\{e^{-\beta  N^{ES}_{0} [\hat{H}_0^{0} ] } \right\} =1$.

\subsection{The sine-Gordon and Thirring models at
$T>0$}\dle{subs:sgth}

We now calculate the SG and MT partition functions through the
following series of steps.

\subsubsection{Removal of UV divergences in the SG model by normal
ordering}\dle{ssubs:div}

Recall that for a scalar field theory in $d$ dimensions with an
interaction of the form $\phi^r$, a diagram with $n$ vertices
and $E$ external
lines has a UV degree of divergence $D$ of \dc{Ryder}
$$
D=d - \left( \frac{d}{2}-1 \right)E + n \left[ \frac{r}{2} (d-2) -d
\right].
$$
Hence with $d=2$,
$D = 2 - 2n$ so that other than the propagator (\dr{bosprop}), 
the only divergent diagrams $\forall r$ are tadpole diagrams.  
Thus, as we now show, all UV divergences are removed by thermal
normal ordering the SG Hamiltonian
\ba
\hat{H}_{SG} &=& \int_0^{L} dx \left[ \frac{\hat{\pi}^2}{2} +
\frac{1}{2}
\left(
\frac{\paa
\hat{\phi}}{\paa x} \right)^2 -
\frac{\alpha_0}{\lambda^2} \cos \lambda \hat{\phi} - \gamma_0 \right]
\nonumber
\\
&=:& \hat{H}_0  - \int_0^{L} dx \left[ 
\frac{\alpha_0}{\lambda^2} \cos \lambda \hat{\phi} + \gamma_0 \right].
\dla{HSG}
\ea
To carry out this procedure, the Hamiltonian must be divided into a free
and an interacting part.  
Although the term $\cos \lambda \hat{\phi}$ itself contains a
mass term on expansion in powers of $\lambda$, we want to
keep $\lambda$ of arbitrary size.  Consider therefore
$$
\hat{H}_{SG} = \left[ \hat{H}_0 +
\int_0^{L} dx \left( \half \rho^2 \hat{\phi}^2 \right) \right] 
- \left[ \int_0^{L} dx \left( 
\frac{\alpha_0}{\lambda^2} \cos \lambda \hat{\phi} + \half \rho^2
\hat{\phi}^2 + \gamma_0 \right)
\right],
$$
so that perturbations are about a scalar field of mass
$\rho$.  To take account of this fact, TNO is now denoted by
$N^{ES}_{\rho} [\bullet]$ and similarly we add an extra mass label to 
 the propagators  (\dr{bosprop}):
%;  $\Delta_{T}(x) \rightarrow
%\Delta_T(x;\tilde{\mu})$\dnote{But in (\dr{bosprop}) the scale 
%$\tilde\mu$ has already been introduced}. 
%Hence in the case of $\hat{H}_{SG}$ above
hence we will be dealing here with $\Delta_T(x;\rho)$.  The link between $\rho$
and the corresponding regularisation scale in path-integral methods is
discussed in section \dr{sec:pi}.

TNO of (\dr{HSG}) may be carried out by using the
identity\footnote{These equalities may at first seem
surprising, but they should be clarified by recalling that here the
advanced and retarded thermal propagators are equal so that
$\Delta^>_{T}(x-y;\rho) = \Delta_T(x-y;\rho) = \left[
{\hat{\phi}}^+(x),{\hat{\phi}}^-(y)\right]$ where the positive and
negative parts are given in (\dr{phipl})-(\dr{phiminus}).}
\ba
e^{i \int_c d^2x j(x) \hat{\phi}(x)} &=& N^{ES}_{\rho} 
\left[ e^{i \int_c d^2x j(x)
\hat{\phi}(x)} \right] e^{\frac{1}{2} \int_c d^2x 
\int_c d^2y j(x)
\Delta_T(x-y;\rho) j(y)}
\nonumber
\\
&=& 
T_c \left[ e^{i \int_c d^2x
j(x) \hat{\phi}(x)} \right].
\dla{identity1}
\ea
Notice that had we used $T=0$ normal ordering, the zero-temperature
propagator would have appeared in (\dr{identity1}) rather than the
finite temperature one:  TNO means that the $Q$ variables of
(\dr{qs}) are built in from the start.  As in
\dc{Coleman} we regulate the  UV divergence of
$\Delta_T(x;\rho)$ by cutting off the theory and replace
$\Delta_T(x;\rho)$ by
$$
\Delta_{T}(x;\rho;\Lambda) = \Delta_{T}(x;\rho) -
\Delta_{T}(x;\Lambda)
$$
where $\Lambda$ is a large mass. 
Note that this operation cancels
the additive constant $K$ in the propagator (\dr{bosprop}).
Now $\Delta_{T}(x;\rho;\Lambda)$ is both non-singular as well
as $\beta$ independent for $x \rightarrow 0$:
$
\Delta_{T}(0;\rho;\Lambda)  = -\frac{1}{4\pi} \ln 
\left( {\rho^2 /
}{\Lambda^2} \right).
$ 
Also equation (\dr{identity1}) with $j(x) = \lambda \delta (x-y)$
gives
\be
e^{i \lambda \hat{\phi}(y)} = \left( \frac{\rho^2
}{\Lambda^2} \right)^{\frac{\lambda^2}{8\pi} } N^{ES}_{\rho} \left[ e^{i
\lambda \hat{\phi}(y)} \right]
\dle{noexpphi}
\ee
so that if we define
\be
\alpha = \alpha_0  \left( \frac{\rho^2
}{\Lambda^2} \right)^{\frac{\lambda^2}{8\pi} }  
\dle{alpren1}
\ee
then the normal ordered form of the SG potential is just as at
$T=0$ \dc{Coleman};
%\dnote{Say how last expression in
%(\dr{alpren1})  is identical to Angels one}
\be
\frac{\alpha_0}{\lambda^2} \cos \lambda \hat{\phi} =
\frac{\alpha}{\lambda^2}
  N^{ES}_{\rho}
\left[ \cos \lambda \hat{\phi}  \right].
\dle{alpharenorm}
\ee

Also TNO of $\hat{H}_0$ (which is defined in (\dr{HSG})
and is only equal to (\dr{Hofermdef})  for zero mass 
$\tilde{\mu} = 0$) gives
\be
\hat{H}_0 = N^{ES}_{\rho} [\hat{H}_0 ] + E_T(\rho)
\dle{normalH0}
\ee
where, using (\dr{phipl})-(\dr{phiminus}),
\ba
E_T(\rho) &=& \frac{1}{2} \left\{ \left[\hat{\pi}^+,\hat{\pi}^-
\right] + \left[(\paa_0 \hat{\phi})^+,(\paa_0 \hat{\phi})^-
\right] \right\}
\nonumber
\\
&=&
 \sum_k \frac{1}{4} (1 + 2N_{k,\rho}) \frac{2k^2 +
\rho^2}{\omega_{k,\rho}} = E_0(\rho) +  \sum_k \frac{1}{2} N_{k,\rho}
\left( \frac{2k^2 +
\rho^2}{\omega_{k,\rho}} \right),
\nonumber
\ea
and $ E_0(\rho)$ is infinite whereas the $T$-dependent part is
finite. 

Thus combining (\dr{alpharenorm}) with (\dr{normalH0}) gives
$$
\hat{H}_{SG} =  N^{ES}_{\rho} \left[ \hat{H}_0 -
\frac{\alpha}{\lambda^2}\int_0^{L} dx
\cos \lambda \hat{\phi}  - L \gamma \right]
$$
where
\be
\gamma = \gamma_0 - E_T(\rho).
\dle{gam}
\ee
Thus $\alpha_0$ has been multiplicatively renormalised,
$\gamma_0$ has been renormalised by an infinite temperature independent
part appearing in $ E_T(\rho)$ as well as having been shifted by a
finite temperature dependent amount, and 
$\lambda$ is unchanged.

Thermal normal ordering the SG Hamiltonian has therefore absorbed
all UV
infinities just as zero temperature normal ordering does \dc{Coleman},
but it has also introduced some extra $T$-dependent finite terms. 
Its power will
become apparent in the next section when thermal expectation values are
calculated.

\subsubsection{Perturbation theory in the SG model}\dle{ssubs:pt}

We now remove the IR divergences of the thermal boson propagator
 by introducing a mass $\tilde{\mu}$, as in (\dr{bosprop}), 
 into the SG Hamiltonian
(\dr{HSG}).  At the end of the calculation we take $\tilde{\mu}
\rightarrow 0^+$ and hence are free to add the
extra mass term within the normal ordering giving the
Hamiltonian
\ba
\hat{H} &=&  N^{ES}_{\rho} \left[ \hat{H}_0  + \int_0^L dx \left(
\frac{1}{2}
\tilde{\mu}^2
\hat{\phi}^2 -
\frac{\alpha}{\lambda^2}\cos \lambda \hat{\phi} \right) - L \gamma
\right]
\nonumber
\\
&=& N^{ES}_{\rho} \left[ \left( \hat{H}_0^{\tilde{\mu}} - \gamma L \right)
 -
\frac{\alpha}{\lambda^2} 
\int_0^L dx \left( \cos \lambda \hat{\phi} \right)   \right]
\nonumber
\\
&=& N^{ES}_{\rho} \left[ \hat{{\cal{H}}}_0^{\tilde{\mu}}
 -
\frac{\alpha}{\lambda^2}\int_0^L dx \left(  \cos \lambda \hat{\phi}
\right)  
 \right]
\nonumber
\\
&\equiv& \hat{A}_0 + \hat{A}_{I}.
\nonumber
\ea
Here $ \hat{A}_0$ and $\hat{A}_{I}$ denote the free ($\alpha=0$) 
and interacting
Hamiltonians respectively.

As usual in perturbation theory, the thermal
expectation value of Heisenberg operators $\bullet_H$ are written in
terms of interaction picture operators $\bullet$ as 
$$
\ll \bullet_H \gg \; = \frac{\Tr \left\{ e^{-\beta \hat{H}} \bullet
\right\}} {\Tr \left\{ e^{-\beta \hat{H}} \right\}} = \frac{\Tr \left\{
e^{-\beta \hat{A}_0}  T_c
\left[ \hat{U}(t_i - i \beta,t_i) \bullet \right]
\right\}} {\Tr \left\{ e^{-\beta \hat{A}_0} T_c\left[ \hat{U}(t_i - i
\beta,t_i) 
\right] \right\}}
$$
 with times lying on the  contour $C$.  For $t_i = 0$
$$
\hat{U}( - i \beta,0) = T_c \left[ e^{-i \int_{0}^{-i \beta} dt''
\hat{A}_I(t'')}
\right].
$$
Thus on expansion in $\hat{A}_I$
 and working in Euclidean space, the SG partition
function is given by
\be
Z_{SG}(T) 
%= \lim_{\tilde{\mu} \rightarrow 0}
%\Tr \left\{ e^{-\beta \hat{H}} \right\} 
= \lim_{\tilde{\mu} \rightarrow
0} \sum_n  \frac{1}{n!} 
\int_{0}^{\beta} dt_1 \ldots \int_{0}^{ \beta} dt_n  \Tr \left\{
e^{-\beta \hat{A}_0} T_c \left[ \hat{A}_I(t_1) \ldots  \hat{A}_I(t_n)
\right]
\right\},
\dle{partz}
\ee
and one is left to calculate free expectation values;
\be
\ll \bullet \gg'_0 \; = \frac{ \Tr \left\{ e^{- \beta \hat{A}_0} \bullet
\right\} }{  \Tr \left\{ e^{- \beta \hat{A}_0}\right\} }
= \frac{ \Tr \left\{ e^{- \beta N_{\rho}^{ES} \left[
\hat{{\cal{H}}}_0^{\tilde{\mu}} \right] }
\bullet 
\right\} }{  \Tr \left\{ e^{- \beta  N_{\rho}^{ES} \left[
\hat{{\cal{H}}}_0^{\tilde{\mu}} \right] }\right\} }
= \frac{ \Tr \left\{ e^{- \beta N_{\rho}^{ES} \left[ \hat{H}_0^{\tilde{\mu}}
\right] }
\bullet
\right\} }{  \Tr \left\{ e^{- \beta  N_{\rho}^{ES} \left[
\hat{H}_0^{\tilde{\mu}} \right] }\right\} }.
\dle{exo}
\ee
Here the prime label occurs because the operator appearing in the
thermal weight is
$ N_{\rho}^{ES} \left[ \hat{H}_0^{\tilde{\mu}} \right]$, and {\em not}
the simple diagonal operator $ :\hat{H}_0^{\tilde{\mu}}:_{\tilde{\mu}}
= \sum_k \omega_{k,\tilde{\mu}}
\hat{a}^{\dagger}_{k,\tilde{\mu}} \hat{a}_{k,\tilde{\mu}}$ 
as in (\dr{defES}).  Since
the results of \dc{ES} can only be used with this latter thermal
weight, we need to manipulate $\ll\gg'_0$.  For that purpose, 
 note that
$$
\hat{H}_0^{\tilde{\mu}} = N_{\rho}^{ES} 
\left[ \hat{H}_0^{\tilde{\mu}} \right] +
\sum_k \frac{1}{4} (1 + 2 N_{k,\rho}) \left( \frac{
2k^2 + \rho^2 + \tilde{\mu}^2 }{\omega_{k,\rho}} \right)
$$
so that from (\dr{H0}) one has
$$
 N_{\rho}^{ES} \left[ \hat{H}_0^{\tilde{\mu}} \right] = \;
:\hat{H}_0^{\tilde{\mu}}:_{\tilde{\mu}} + 
\sum_k \frac{\omega_{k,\tilde{\mu}}}{2} - 
\sum_k \frac{1}{4} (1 + 2 N_{k,\rho}) \left( \frac{
2k^2 + \rho^2 + \tilde{\mu}^2 }{\omega_{k,\rho}} \right).
$$
Hence 
substituting into (\dr{exo}) gives
$$
\ll \bullet \gg'_0 \;  = \frac{ \Tr \left\{ e^{-
\beta : \hat{H}_0^{\tilde{\mu}}:_{\tilde{\mu}} } \bullet
\right\} }{  \Tr \left\{ e^{- \beta 
: \hat{H}_0^{\tilde{\mu}}:_{\tilde{\mu}}}\right\} } = \; \ll \bullet
\gg_0 
$$
and the results of \dc{ES} may then be used to say that
$
\ll N_{\tilde{\mu}}^{ES} \left[ \bullet \right] \gg_0 = 0
$
whenever  $ \bullet$ is a product of operators (not including a
constant).  
% This property 
%is the main advantage 
%of TNO will be
%used extensively below.
%
Finally, before proceeding, we note from equation (\dr{noexpphi}) that
\be
N^{ES}_{\rho} \left[ e^{i \lambda \hat{\phi}} \right] = \left(
\frac{\tilde{\mu}^2}{\rho^2}
\right)^{\frac{\lambda^2}{8\pi}} N^{ES}_{\tilde{\mu}}  \left[  e^{i 
\lambda
\hat{\phi}} \right] .
\dle{relnexp}
\ee

Following  the steps in \dc{Coleman} for $T=0$, 
let us calculate at $T>0$ the 
free correlator
\ba
\ll T_c \prod_{j=1}^n N^{ES}_{\rho} \left[  e^{i \lambda_j
\hat{\phi}(x_j)}
\right] 
\gg_0 \; &=& \frac{ \Tr \left\{   e^{-\beta N_{\rho}^{ES} \left[
\tilde{H}_0^{\tilde{\mu}} \right] } T_c
\left[ 
 \prod_{j=1}^n N^{ES}_{\rho} \left[ e^{i
\lambda_j \hat{\phi}(x_j)} \right] \right]  \right\} }
{ \Tr \left\{   e^{-\beta N_{\rho}^{ES} \left[
\tilde{H}_0^{\tilde{\mu}} \right]  } \right\} }
\nonumber
\\
& =&  \left( \frac{\tilde{\mu}^2}{\rho^2}
\right)^{\frac{\sum_j \lambda_j^2}{8\pi}} 
\frac{ \Tr \left\{   
e^{-\beta :H_0^{\tilde{\mu}}:_{\tilde{\mu}} } T_c \left[  \prod_{j=1}^n
N^{ES}_{\tilde{\mu}} \left[ e^{i  \lambda_j \hat{\phi}(x_j)} \right] \right] 
 \right\} }
{ \Tr \left\{   
e^{-\beta :H_0^{\tilde{\mu}}:_{\tilde{\mu}} }  \right\} }
\nonumber
\ea
where we have used (\dr{relnexp}). Note that the above correlator is the 
 one we would need to prove the equivalence (\dr{equivoper1}) by 
 expanding in $\alpha$ \dc{Coleman}. 
Also from (\dr{identity1}) with  $j(x) = \lambda_j \delta(x-x_j)$ 
and using the fact that time ordering always puts the operators in a
given order which depends only on their time arguments, one has
$$
T_c \left[ \prod_{j=1}^{n} N_{\tilde{\mu}}^{ES} 
\left[ e^{i  \lambda_j \hat{\phi}(x_j)}
\right] \right] = 
 T_c \left[ e^{i  \sum_j (\lambda_j \hat{\phi}(x_j))} \right]
e^{\frac{1}{2}\Delta_{T}(0;\tilde{\mu}) \sum_j \lambda_j^2 }.
$$
Hence
\ba
T_c \left[ \prod_{j=1}^{n} N_{\tilde{\mu}}^{ES} 
\left[ e^{i  \lambda_j \hat{\phi}(x_j)}
\right] \right] &=& 
N^{ES}_{\tilde{\mu}} \left[ e^{i  \sum_j (\lambda_j \hat{\phi}(x_j))} \right]
e^{-\frac{1}{2} \sum_j \sum_k \lambda_j \Delta_{T}(x_j - x_k;\tilde{\mu})
 \lambda_k}
e^{\frac{1}{2} \sum_j \lambda_j^2 \Delta_{T}(0;\tilde{\mu})}
\nonumber
\\
&=&
N^{ES}_{\tilde{\mu}} \left[ e^{i  \sum_j (\lambda_j \hat{\phi}(x_j))} \right]
e^{-\sum_{j>k}  \lambda_j \Delta_{T}(x_j - x_k;\tilde{\mu}) \lambda_k}
\nonumber
\\
& = &
N^{ES}_{\tilde{\mu}} \left[ e^{i  \sum_j (\lambda_j \hat{\phi}(x_j))} \right]
\prodj\prod_{j>k} \left[ \beta^2 \tilde{\mu}^2 
|Q(x_j - x_k)|^2 \right]^{\frac{\lambda_j\lambda_k}{4 \pi}},
\nonumber
\ea
giving, finally,
\ba
\ll T_c \left[ \prod_{j=1}^{n} N_{\rho}^{ES} 
\left[ e^{i  \lambda_j \hat{\phi}(x_j)}\right] \right] \gg_0
&=& 
\left( \frac{\tilde{\mu}^2}{\rho^2}
\right)^{\frac{\sum_j \lambda_j^2}{8\pi}}  
\ll N^{ES}_{\tilde{\mu}} \left[ e^{i 
\sum_j (\lambda_j \hat{\phi}(x_j))} \right] \gg_0 \times
\nonumber
\\
&&\; \; \; \; \; \; \; \; \; \; 
\times \prodj\prod_{j>k} \left[ \beta^2 
\tilde{\mu}^2 |Q(x_j - x_k)|^2 \right]^{\frac{\lambda_j
\lambda_k}{4 \pi}}
\nonumber
\\
& = & 
\left( \frac{\tilde{\mu}^2}{\rho^2}
\right)^{\frac{1}{8\pi}\sum_j \lambda_j^2}  \prodj\prod_{j>k} \left[ \beta^2
\tilde{\mu}^2 |Q(x_j - x_k)|^2 \right]^{\frac{\lambda_j
\lambda_k}{4 \pi}}
\dla{answer}
\ea
where the last line follows
from expanding the exponential inside the normal ordered term and
noting that all terms which contain
$\hat{\phi}^n$ vanish (by (\dr{NOcond})) apart from the one for which
$n=0$, giving a contribution of 1. It is at this stage that thermal
normal ordering has been crucial for the calculation---had we used
zero temperature normal ordering, this correlator would have been very
much more difficult to calculate.  Note that the terms proportional
to
$\tilde{\mu}$ in (\dr{answer}) have a contribution
\be
(\tilde{\mu}^{2})^{\frac{\sum_n \lambda_j^2}{8 \pi}} \prodj\prod_{j>k}
(\tilde{\mu}^{2})^{\frac{ \lambda_k \lambda_j}{4 \pi}} = 
\tilde{\mu}^{\frac{\left(\sum_n\lambda_j \right)^2}{4 \pi}}
\label{contri}
\ee
so unless $\sum_n
\lambda_j = 0$, the result vanishes as $\tilde{\mu} \rightarrow 0$, the limit
in which we are interested.  Hence we only consider configurations with 
$\sum_n
\lambda_j = 0$, a condition which will become analogous to the
fermion chiral selection rule of section \dr{sec:pi}.  In
(\dr{answer}) $n$ must therefore be even.  If we let $m=n/2$,
$y_j=x_j$ for $j=n/2+1,\dots,n$ and define
$ \hat{A}_{\pm} = N^{ES}_{\rho} \left[ e^{\pm i
\lambda \hat{\phi}} \right]$ then
\be
\ll T_c \prod_{j=1}^{m}  \hat{A}_+(x_j)\hat{A}_-(y_j) \gg_0 \; =
\prod_{j =1 }^m \frac{ \prod_{j > k}^{m} \left[ \beta^4
\rho^4 |Q(x_j - x_k)|^2 |Q(y_j - y_k)|^2
\right]^{\frac{\lambda^2}{4\pi}} }{
 \prod_{k=1}^m \left[ \beta^2
\rho^2 |Q(x_j - y_k)|^2 \right]^{\frac{\lambda^2}{4\pi}} }
\dle{Acorrel}.
\ee
%\dnote{Note that for $m=1$ the answer is
%NOT zero!  The top might appear to be zero, but is in fact 1 ,and the
%bottom gives an answer}
To calculate the partition function we need
\ba
\lefteqn{ \ll T_c \prod_{j=1}^{2n}  N^{ES}_{\rho}  \left[ \cos \lambda
\hat{\phi}(x_j) \right]\gg_0 \; = }
\nonumber
\\
&&
\frac{1}{2^{2n}} \frac{(2n)!}{(n!)^2} \prod_{j =1 }^n
\frac{ \prod_{j > k}^{n} \left[ \beta^4
\rho^4 |Q(x_j - x_k)|^2 |Q(y_j - y_k)|^2
\right]^{\frac{\lambda^2}{4\pi}} }{
 \prod_{k=1}^n \left[ \beta^2
\rho^2 |Q(x_j - y_k)|^2 \right]^{\frac{\lambda^2}{4\pi}} }
\nonumber
\ea
which follows from (\dr{Acorrel}) and (\dr{contri}).  
Observe that the total number of
$\beta
\rho $ factors in this expression is
$
(\beta \rho)^{- \frac{\lambda^2}{4\pi} (2n)}$.  Hence the partition
function (\dr{partz}) is 
\ba
\lefteqn{ Z_{SG}(T) = \Tr \left\{ e^{-\beta \hat{H}_{SG}} \right\} = } 
\nonumber
\\
&=&
Z_0^{B}(T) \sum_n \frac{1}{(2n)!}  \left( \frac{-\alpha}{ \lambda^2}
\right)^{2n}
\int_{T} d^2x_1 \ldots \int_{T} d^2x_{2n}  
\ll
 T_{c}\left[ 
\prod_{j=1}^{2n} N_{\rho}^{ES} \left[ \cos \lambda \hat{\phi}(t_j)\right]
\right] \gg_0 
\nonumber
\\
&=&
Z_0^{B}(T)
 \sum_n  \left( \frac{1}{n!} \right)^2  \left[
\frac{\alpha}{2\lambda^2}
\left( \frac{T}{\rho} \right)^{\frac{\lambda^2}{4 \pi}} \right]^{2n}
\nonumber
\\
 && \; \; \; \; \times  \prod_{j=1}^n \int_{T} d^2x_j  \int_{T} d^2y_j 
\frac{ \prod_{j > k}^{n} \left[ |Q(x_j - x_k)|^2 |Q(y_j - y_k)|^2
\right]^{\frac{\lambda^2}{4\pi}} }{ 
 \prod_{k=1}^n \left[  |Q(x_j - y_k)|^2
\right]^{\frac{\lambda^2}{4\pi}} }.
\dla{sgparfun}
\ea
where $Z_0^{B}(T) =  \Tr \left\{  
e^{-\beta \hat{A}_0} \right\} = e^{\beta \gamma_0 L } Z_0^{0,B}(T)$ and it is
finite for $\gamma_0$ satisfying (\dr{gam}) with $\gamma$ finite.

One might wonder whether there are any extra divergences in the 
$x_j$, $y_j$ integrals in (\dr{sgparfun}), which, for a given $n$
 correspond to closed diagrams with $2n$ vertices and an arbitrary number of 
 loops. Recalling the asymptotic behaviour of the 
$Q$ variables in (\dr{asbeq}), one realises that the integrals
in (\dr{sgparfun}) are always finite  for large spatial $x_j$ or
$y_j$, because the number of $Q^2(x_j)$ powers  in the denominator is
always bigger than that in the numerator.  As for the behaviour of the
integrand when $x_j\rightarrow 0$ we see that the integral converges
provided $\lambda^2<4\pi$.  Hence in this case, all the UV
divergences of the theory are removed by renormalising $\alpha$. 
 If $\lambda^2>4\pi$, new divergences could appear in principle. However, 
 it can be shown that as long as $\lambda^2<8\pi$, those extra divergences can
 also be removed, by subtractions to the free energy \cite{agg80}. The 
 same argument holds in our case, since the UV divergences are $T$ 
 independent. In fact, for $\lambda^2<8\pi$ the theory is super-renormalisable 
 \dc{Coleman}. For $\lambda^2\geq 8\pi$ though, the theory is no longer 
 super-renormalisable and indeed the ground-state disappears from the 
 spectrum  \dc{Coleman}.  These bounds should borne in mind and, 
although formally our expressions and results hold  for any 
$\lambda$, they will remain finite only for $\lambda^2<8\pi$ where we will
 assume that the  $T=0$ subtractions have been already  made.
%\dnote{I WOULD CUT
%THIS: Notice that this 
% does not alter our arguments, since we are ignoring 
% $T$ independent terms in the  free energy.}

\subsubsection{Perturbation theory in the massive Thirring
Model}\dle{ssubs:pft}

In the Thirring model, the situation is somewhat unusual as 
perturbation theory not about the zero coupling $g=0$ theory, but about
the massless $m_0 = 0$ theory.  The correlator corresponding to
(\dr{Acorrel}) is therefore
\be
\ll T_c \prod_{j=1}^{n}   \hat{\sigma}^R_+(x_j) \hat{\sigma}^R_-(y_j) 
\gg_{\displaystyle\tiny\begin{array}{c}m=0\\g\neq
0\end{array}} 
\dle{sigcorr}
\ee
where $R$ stands for renormalisation
%\dnote{Cut????:  The
%renormalisation condition  (see below) is necessary to deal with the
%product of two operators at  the same space-time point.} 
and
$$
\hat{\sigma}_{\pm}(x) =  \overline{\hat{\psi}}(x) P_\pm
\hat{\psi}(x) \quad \mbox{with} \quad P_\pm=\half (1 \pm \gamma_5).
$$
At $T=0$ when the thermal expectation values are replaced by vacuum
expectation values, such correlators have been calculated in a
canonical operator approach by Klaiber \dc{Klaib}. 
His techniques have been used extensively by Coleman \dc{Coleman} as well
as by others for example in the study of the Schwinger model
\dc{sch62}. 
However, the generalisation of Klaiber's work to $T>0$ does
not seem to have been carried out.  As in the case of the SG
model, this is obtained by firstly replacing zero temperature 
normal ordering by thermal normal ordering and then using thermal
expectation values rather than the vacuum expectation values of \dc{Klaib}.
This changes all zero temperature propagators to
their finite temperature analogues just as in section
\dr{ssubs:tno}, and is equivalent to replacing
$x_{\alpha} \rightarrow \beta Q_{\alpha}$.  Furthermore, the
regularisation remains unchanged from the $T=0$ case---that is,
given the parameter $\delta$ defined in \dc{Klaib};
$$
\delta = \frac{g^2}{4 \pi} \frac{2\pi + g^2}{\pi + g^2},
$$
it follows from equations (VII.5), (VII.4) and (VII.1) of \dc{Klaib}
that in order for the correlator (\dr{sigcorr}) of $\hat{\sigma}_{\pm}(x)$'s
to be finite, these operators must be regularised as
\be
\hat\sigma_{+}^R(x) =
%\lim_{x \rightarrow y} [M (x-y)]^{2 \delta}
%\psi_1^{\dagger}(x) \psi_2(y)
%\nonumber
%\\
  \lim_{\varepsilon' \rightarrow 0} [M \varepsilon']^{2 \delta}
\hat{\psi}_1^{\dagger}(x) \hat{\psi}_2(x-\varepsilon').
\dle{Sigrenc}
\ee
Here the $\hat{\psi}_i$'s label the components of the spinor,
$\varepsilon'$ is a 2-vector, and $M$ in is
an arbitrary renormalisation scale which, as opposed to
\dc{Coleman}, has been included so as not to 
alter the dimension of the operators.  
Finally, $\hat{\sigma}_-^R$ is given by the
hermitian conjugate of (\dr{Sigrenc}).  As we will verify also in
section \dr{piparfun}, following these steps and 
using the equations of \dc{Klaib} mentioned above, one obtains
\ba
\lefteqn{ \ll T_c \prod_{j=1}^{n}   \hat{\sigma}_+^R(x_j)
\hat{\sigma}_-^R(y_j) \gg_{\displaystyle\tiny\begin{array}{c}m=0\\g\neq
0\end{array}}  \; = } 
\nonumber
\\
&& 
\left( \frac{M}{2}\right)^{2n} \prod_{j=1}^{n} \frac{ 
\prod_{ j>k}^{n}
\left[
\beta^4 M^4 |Q(x_j - x_k)|^2 |Q(y_j - y_k)|^2
\right]^{\frac{\pi}{\pi + g^2}} }{
 \prod_{k=1}^n \left[ \beta^2
M^2 |Q(x_j - y_k)|^2 \right]^{\frac{\pi}{\pi + g^2} } }.
\dla{sigcorrel}
\ea
Observe that up to the factor of $(M/2)^{2n}$, equations
(\dr{sigcorrel}) and (\dr{Acorrel}) are identical at $T>0$ 
as long as the two theories are renormalised at the same scale, $M =
\rho$, and provided that
$$
\frac{\pi}{\pi + g^2} = \frac{\lambda^2}{4 \pi}
$$
which is just (\dr{equivconst1}).
The partition function then follows directly, and is given by
\ba
Z_{MT} (T,\mu = 0) &=& 
Z_0^F (T)\sumn \left(\frac{1}{n!}\right)^2 
\left[\frac{m}{2\beta}\left(\frac{\rho}
{T}\right)^{\kappa^2/\pi}\right]^{2n} \prodjint \times
\nonumber
\\
&& \; \; \; \; \;
\frac{\prod_{k<j}\left[
Q^2(x_j-x_k)Q^2(y_j-y_k)\right]^{1-\kappa^2/\pi}}{
\prod_{k=1}^n\left[Q^2(x_j-y_k)\right]^{1-\kappa^2/\pi}}.
\dla{thparfuna}
\ea
Here $Z_0^F$ is given in (\dr{partferm}), 
\be
\kappa^2 = \frac{g^2}{1+g^2/\pi}
\dle{kapdef}
\ee
and $m$ is the renormalised mass (see also section \dr{sec:pi})
\be
m = m_0 (\varepsilon' \rho)^{-2 \delta}.
\dle{mren}
\ee
In perturbation theory the SG and MT model partition functions are 
therefore identical term by term if we identify 
$$
\rho m = \frac{\alpha}{\lambda^2}
$$
as in (\dr{equivconst2}).

In order to prove the relationship between arbitrary products of zero
charge operators at different space-time points in the two theories and 
 treat the $\mu\neq 0$ problem,
we will turn to more powerful path integral methods.

\section{Bosonisation with Path-Integral methods at $T>0$ and
$\mu=0$}
\dle{sec:pi}

Again we work in path-integral imaginary time formalism 
of thermal field theory and hence in Euclidean space-time with metric
(+,+).  Now we change notation slightly so that $t=\bar{t}$ or $x^0$ 
denote the (real) Euclidean time, but, unless stated otherwise, the
remaining notation follows that of \dc{zj}.  As usual the exponential
factor in the path integral is $\exp(-S)=\exp (-\intT {\cal L}(x,t))$
with $\intT\equiv \int_0^\beta dt\int_{-\infty}^{+\infty} dx$, and the
Euclidean Lagrangian densities for the SG and massive Thirring models
are respectively 
$$
{\cal L}_{SG}[\phi] = \frac{1}{2} \paa_\mu \phi  \paa^\mu \phi -
\frac{\alpha_0}{\lambda^2} \cos \lambda \phi ,
$$
$$  
{\cal{L}}_{MT} [\bar\psi,\psi] =  -\bar{\psi} (\pabar + m_0)\psi +
\frac{1}{2}g^2  j_\mu (x) j^\mu (x)  
$$   
where we set $\gamma_0 = 0$ in the SG Lagrangian.  The
Euclidean gamma matrices are 
$$
\gamma^0=
\left( \begin{array}{cc} 
   0 & 1 \\   
  1    &   0  
\end{array} \right) \,  \, \, \,    
\gamma^1=\left( \begin{array}{cc}     
    0 & -i \\      
       i    &   0  
\end{array} \right) \,  \, \, \, 
\gamma_5 = -i \gamma^0 \gamma^1 =
\left( \begin{array}{cc}    
          1 & 0 \\        
    0    &   -1    
\end{array} \right)     
$$
so that
$$
\{\gamma^\mu,\gamma^\nu\} = 2\delta^{\mu\nu},
\; \; \; \; \; 
\gamma_\mu\gamma_5 = -i\epsilon_{\mu\nu}\gamma_\nu 
$$
with $\epsilon_{01}=1$.  

Let us briefly sketch the outline of this section. Firstly, in 
section  \dr{subs:pigf}, some useful  free boson and fermion 
correlators are calculated in the path integral
formalism for $T>0$ and $\mu=0$.  Here we make use of the free generating 
functionals and thereby reproduce the results for those correlators 
obtained with canonical methods in the previous section. The aim of 
section \dr{ren} is to recall how renormalisation takes place in the 
path integral and, in turn, to compare with our renormalisation conventions 
in the canonical formalism. Finally, in section \dr{piparfun} we 
re-obtain the equivalence between the partition functions of the two 
models, now from the path integral viewpoint.  This then serves
to introduce the SG and MT generating functionals
which will be needed later to re-derive the correlator
(\dr{sigcorrel}) in this language,
%\dnote{``hence providing support to our conjectures about the 
% $T>0$ extension of Klaiber'' (or something like this, 
%depending on how do you say it before)}, 
as well as to introduce the auxiliary field technique we will be 
using extensively.  Besides, most of the path integral techniques
and results needed in the rest of the paper will also be introduced in 
section \dr{piparfun}.   As in the previous section, we always 
proceed by expanding around the $\alpha_0=m_0=0$ theories. 

\subsection{Free generating functionals and useful
correlators}\dle{subs:pigf}

We start from the free boson Euclidean generating functional 
\ba
Z_0^B[J;T]&=&N_\beta
 \pib \exp\left\{-\left[\intT \frac{1}{2}\left[
(\partial_\alpha \phi)^2+
{\tilde{\mu}}^2\phi^2\right]+J(x)\phi(x)\right]\right\}\nonumber\\
&=&Z_0^B[0;T]\exp\left\{\frac{1}{2}\intT\int_T d^2y \ J(x)
\Delta_T (x-y) J(y)\right\}
\dla{freebosgenfun}
\ea
where the propagator is given in (\dr{bosprop}) for small 
$\tilde{\mu}\beta$ and the free boson
partition function is $Z_0^B(T) = Z_0^B[0;T]$ as in
(\dr{partbos}).  Note that from this section onwards the $\tilde{\mu}$
labels in propagators and partition functions will be removed
since
%\dnote{CUT?: we are only interested in the $\beta\tilde{\mu}\ll
%1$ limit and } 
there is no possible confusion with any other mass
scale. 
Similarly the free massless fermion generating functional is
\ba
Z^F[\bar\eta,\eta;T]&=&N_\beta^F
\pif\exp\left[\intT\left(\bar\psi\pabar\psi+\bar\eta\psi
+\bar\psi\eta\right)\right]\nonumber\\
&=&Z^F[0,0]\exp\left[-\intT\int_T d^2y \bar\eta (x)S(x-y)\eta(y)\right]
\dla{fergenfun}
\ea
where $Z^F[0,0] = Z_0^F$ in (\dr{partferm}) and $N_\beta^F$ is the fermionic 
 counterpart of $N_\beta$ in (\dr{thav}).  The free massless
propagator $S(x-y)$ can also be expressed in terms of the $Q$
variables as \dc{Belg}
\be
S_{\alpha\beta} (x) = 
-\frac{1}{2\beta}\frac{\Qbar_{\alpha\beta} (x)}{Q^2(x)}
\dle{fermprop}
\ee
where the indices $\alpha$, $\beta$ are Dirac indices.

The path-integral counterpart of the free correlator (\dr{Acorrel}) is easily
calculated within the path-integral formalism.  We outline the main
steps as they are typical of those required below.  
 We define now $A_\pm=\exp\left[\pm i\lambda\phi\right]$ 
(remember that there is
no operator normal ordering in the path integral), so that, 

$$
\ll T_c\prodj A_+(x_j) A_- (y_j)\gg_0=
%\exp\left[i\lambda\left(\phi(x_j)-\phi(y_j)\right)\right]
\frac{Z_0^B[\tilde J;T]}{Z^B_0(T)}
$$
with
$$
\tilde J(z)=-i\lambda\sum_{j=1}^n\left[\delta^{(2)}(z-x_j)
-\delta^{(2)}(z-y_j)\right].
$$
 Unlike  section \dr{sec:op}, here the UV
divergence of the propagator (\dr{bosprop}) will be  regulated by replacing
\be
 Q^2(0,0)\rightarrow  Q^2(\varepsilon_0,\varepsilon_1)=
 T^2 \varepsilon^2+\Od(\varepsilon^3)
\dle{bosonreg}
\ee 
 where $\varepsilon_\alpha \rightarrow 0^+$ and 
$\varepsilon^2=\pi^2(\varepsilon_0^2+\varepsilon_1^2)$. 
Then from (\dr{freebosgenfun}),  (\dr{bosprop}) and 
(\dr{bosonreg}) it follows that
\be
\ll T_c \prodj A_+(x_j) A_- (y_j)\gg_0=
( T\varepsilon)^{n \lambda^2 /2\pi} \prod_{j=1}^n \frac{\prod_{k<j}\left[
Q^2(x_j-x_k)Q^2(y_j-y_k)\right]^{\lambda^2/4\pi}}
{\prod_{k=1}^n\left[Q^2(x_j-y_k)\right]^{\lambda^2/4\pi}}.
\dle{bosfreecorr}
\ee
Observe that this correlator is ill-defined as $\varepsilon\rightarrow 0^+$.
 Since we are still dealing with the free theory, the parameter $\lambda$  
 entering the definition of the $A_\pm$ operators above is, so far, 
 arbitrary and, hence, the above correlator could, for instance,  diverge
 or vanish by taking $\lambda$ purely imaginary or real respectively. 
 This behaviour is due to the
short-distance (UV) divergent behaviour of the composite operator
$\exp[i\lambda\phi(x)]$ which needs then to be renormalised (see section
\dr{ss:sgren} below).  Also note that the
$\tilde{\mu}$ dependence in (\dr{bosfreecorr}) has vanished for the
same reasons as those of
 section \dr{ssubs:pt}, i.e,  because we have chosen equal
numbers of $A_+$ and $A_-$. It is not difficult to see that, had we
chosen  different numbers, the above correlator would have
been proportional to $\tilde{\mu}^\alpha$ with $\alpha>0$ and hence 
it would vanish in the $\tilde{\mu}\rightarrow 0^+$ limit.  This is
the bosonic counterpart of the so called fermion chiral selection
rule which we will discuss below. 

Similarly we need to calculate the fermion free correlator of an 
arbitrary number of insertions of the operators 
$\sigpm =\bar\psi (x)P_\pm \psi (x)$.  To do that, recall that the
massless fermion theory is invariant under chiral transformations 
$\psi\rightarrow\exp(i\alpha\gamma_5)\psi$ with $\alpha$ real arbitrary. 
Under such transformation 
$\sigpm\rightarrow\exp (\pm 2i\alpha)\sigpm$ and therefore the
thermal average of a product of $\sigpm$ operators will vanish in 
the massless case unless the number of $\sigma_+$ and $\sigma_-$ is
the same. This is the  chiral selection rule, which only holds for
$m_0=0$.  Following \dc{zj}, the required correlator is obtained by
shifting $\bar\psi\rightarrow \bar\psi\gamma^0$ so
that, naming $\psi_a$ with $a=1,2$  the two components of the bispinor
$\psi$, the free massless theory now decouples into two free
theories for the spinors $\psi_a$, with propagators
\be
S_a (x-y)=\fouri \frac{ip_0\pm p}{p^2}=-\frac{1}{2\beta}
\frac{1}{Z_Q^\pm (x-y)}.
\dle{freeferprop12}
\ee
The $+,-$ signs correspond to $a=1,2$ respectively and 
$Z_Q^\pm$ are given in (\dr{zq}). 
Now the transformation $\bar\psi\rightarrow \bar\psi\gamma^0$ takes
$\sigma_+\rightarrow \bar\psi_2\psi_1$ and 
$\sigma_-\rightarrow \bar\psi_1\psi_2$. Recall also the thermal 
averaged version of Wick's theorem for fermions \dc{zj} which is
\be
\ll  T_c  \prodj \bar\psi_a (x_j)\psi_a (y_j)\gg_0 =\det S_a
\dle{wickfer}
\ee
where the subscript 0 for fermion correlators will denote the 
free massless case $m_0=g=0$ and 
 $S_a$ denotes the $n\times n$ matrix with elements 
$(S_a)_{jk}=S_a (x_j-y_k)$. Hence, using the result \dc{zj,Belg}
\be
(-1)^{n+1}\det\frac{1}{Z_Q^\pm (x_j-y_k)}=
\prodj\frac{\prod_{k<j}
Z_Q^\pm(x_j-x_k)Z_Q^\pm (y_j-y_k)}
{\prod_{k=1}^n Z_Q^\pm(x_j-y_k)}
\dle{det}
\ee
where $Z_Q^\pm$ is given  in (\dr{zq})
we have
\be
\ll  T_c  \prodj \sigma_+ (x_j) \sigma_- (y_j) \gg_0 =
 (2\beta)^{-2n}\prodj
\frac{\prod_{k<j}\left[
Q^2(x_j-x_k)Q^2(y_j-y_k)\right]}
{\prod_{k=1}^n\left[Q^2(x_j-y_k)\right]}.
\dle{ferfreem=0corr}
\ee
 Notice that the above
correlator has exactly the same structure as the boson correlator
(\dr{bosfreecorr}).  However, (\dr{ferfreem=0corr}) is finite and 
 well-defined, 
unlike (\dr{bosfreecorr}),  since it contains no products of  fields at
the same point and also because there are no mixing terms between 
$\psi_1$ and $\psi_2$ in the Lagrangian. We will come back to this
point in section \dr{renth}.

\subsection{Renormalisation}
\dle{ren}

Here we recall the path integral renormalisation procedure for our 
 present case.  We want to remark that our regularisation and renormalisation 
conventions are going to be the same  as those discussed, for instance, 
 in \dc{zj} for
$T=\mu=0$.  That renormalisation is independent of $T$ and $\mu$ is an 
expected result which will be explicitly checked throughout the rest 
 of the paper. In addition, note that we need only to analyse 
 renormalisation in the free ($\alpha_0=0$) SG model and in the 
massless ($m_0=0$) Thirring model. The 
reason is that, just as with the canonical methods, we treat
the SG model and the massive Thirring model always in terms of free
and $m_0=0$ correlators, with an arbitrary number of insertions of 
$(\alpha_0/\lambda^2)\cos\phi(x)$ and $ m_0\bar\psi\psi (x)$ respectively.

\subsubsection{sine-Gordon model}
\label{ss:sgren}

As commented above, the ill-defined behaviour of
 the free correlator (\dr{bosfreecorr}) is caused by
the short-distance behaviour of the free propagator in
(\dr{bosonreg}). This problem can be solved
 by renormalising the composite operator 
$\exp[ia\phi(x)]$ with $a$ arbitrary as
\be
\left[\exp\left[i a \phi (x)\right]
\right]_{bare}=(\varepsilon \rho)^{a^2/4\pi}\left[
\exp\left[i a \phi(x)\right]\right]^R
\dle{renexp}
\ee
where $\rho$ is an arbitrary renormalisation scale.  Note that this
expression is analogous to (\dr{noexpphi}) with the identification
$$
\varepsilon = \frac{1}{\Lambda},
$$
although in the operator formalism the renormalisation was carried out
through TNO.  Also observe that the dependence with the scale $\rho$ 
is the same, 
though it has slightly different origins---in the canonical operator
approach $\rho$ corresponded to an arbitrary mass at which normal
ordering was performed whereas here there is no normal ordering and $\rho$ 
 is an arbitrary renormalisation scale.

As observed in section \dr{ssubs:pt}, all the infinities that appear
in the theory are just those coming from (\dr{renexp}), after free-energy 
 subtractions,  provided  
$\lambda^2<8\pi$. Those divergences  can be removed  by defining the
renormalised coupling constant $\alpha$ as

\be
\alpha=\alpha_0 (\varepsilon\rho)^{\lambda^2/4\pi}
\dle{renalpha}
\ee
({\em c.f.}\ equation (\dr{alpren1})), and the coupling constant
$\lambda$ does not need renormalisation. We will check explicitly below 
that with this coupling constant  renormalisation all our results remain 
finite (for $\lambda^2<8\pi$, as explained above).

\subsubsection{Thirring model}\dle{renth}

We have seen that the $\sigpm$ correlators are finite in the $m_0=g=0$
case because the two components of the bispinor do not mix in the
Lagrangian. If $g\neq 0$, $m_0=0$, the chiral symmetry is still unbroken so
that $\sigpm$ correlators again appear in the same number.  However, when 
$\bar\psi\rightarrow\bar\psi\gamma^0$, 
$(\bar\psi\gamma^\mu\psi)^2\rightarrow 4\bar\psi_1\psi_1\bar\psi_2\psi_2$
and therefore there is mixing between $\psi_1$ and $\psi_2$. Hence
$$
\ll \bar\psi_2 (x)\psi_1 (x)\bar\psi_1 (y)\psi_2 (y)\gg\neq
 -\ll \bar\psi_2 (x) \psi_2 (y) \gg \ll \bar\psi_1 (y) \psi_1 (x) \gg
$$
when $g\neq 0$, so that products of fields at the same point appear and
then the $\sigpm$ correlator becomes divergent. As for the
boson operator $\exp [ia\phi(x)]$, the $\sigpm$ composite operators need
renormalisation.  Also as in SG, those are the only infinities to
consider and now they can be absorbed in the renormalised
mass $m$.  We anticipate here the results for such renormalisations, which
 we will explicitly check below, 
\ba
\left[\sigpm\right]^R&=&(\varepsilon\rho)^{\kappa^2/\pi}
\left[\sigpm\right]_{bare}
\dla{rensigma}
\\
 m&=&m_0 (\varepsilon\rho)^{-\kappa^2/\pi}
\dle{renbarm}
\ea 
where  both models  are renormalised at the same scale $\rho$ and
$\kappa$ is given in (\dr{kapdef}). Notice that our renormalisation 
conventions here for the fermion case differ from those in section 
 \dr{ssubs:pft}, which are those used in \cite{Klaib} 
(compare for instance (\dr{renbarm}) with (\dr{mren})). For that reason, 
 we have used different notations for our short-distance cut-offs, namely,
 $\varepsilon$ and $\varepsilon'$. However, observe that from  
(\dr{renbarm}) and  (\dr{mren}) and the definitions of $\kappa$ and $\delta$,
  they are related through

$$
\varepsilon'=\varepsilon\left(\varepsilon\rho\right)^{-\frac{g^2}{2\pi+g^2}}
$$
and since the exponent of $\varepsilon$ in the r.h.s of the above equation 
 is positive  for $\lambda^2<8\pi$ according to (\dr{equivconst1}), we can 
 take both $\varepsilon,\varepsilon'\rightarrow 0^+$ consistently. 
In addition, we will see below that 
the answer for the (finite) renormalised correlators and partition 
 functions 
 in both canonical and path integral approaches is the same, even though 
 our renormalisation conventions are different, which is a necessary 
 consistency check.

\subsection{Generating functionals and partition functions}
\dle{piparfun}

\subsubsection{The SG generating functional}

As usual, all correlators can be obtained from the generating
functional which, in the SG model, is
\be
Z_{SG}[J;T]=
N_\beta
 \pib \exp\left\{-\left[\intT {\cal L}_{SG}[\phi]+J(x)\phi(x)\right]\right\}
\dle{sggf}
\ee
where $Z_{SG}(T)\equiv Z_{SG}[0;T]$
%\dnote{You were using $Z_{SG}(T,\mu=0)$
% instead. I think it is better to change yours, since $\mu$ does not appear
% in $Z_{SG}$  
% until section 5 and the notation stays clearer not writing $\mu=0$ 
%everywhere, so that I've changed it in your section} 
is the SG partition function. 
Expanding formally in powers of $\alpha_0/\lambda^2$ gives
$$
Z_{SG}[J;T]=\sumn\frac{1}{n!}
\left(\frac{\alpha_0}{2\lambda^2}\right)^n\sum_{k=0}^n 
 \nk \prodj \intT d^2 x_j \ 
 Z_0^B[J+\bar J;T]
$$
with
$$
\bar J (z)=-i\lambda\sumj s_j \delta^{(2)} (z-x_j)\quad ; \quad
s_j=\left\{\begin{array}{l} + \qquad   j=1,...,k\\
- \qquad   j=k+1,...,n \end{array}\right. .
$$ 
Hence, from (\dr{freebosgenfun}) we have
\ba
Z_{SG}[J;T]&=&Z_0^B[J;T]\sumn \frac{1}{n!}
\left(\frac{\alpha_0}{2\lambda^2}\right)^n
\sum_{k=0}^n 
 \nk\prodj \intT d^2 x_j
\exp\left\{-\frac{\lambda^2}{2}n\Delta_T (0)\right.
\nonumber\\
&-&\left.i\lambda\sumj s_j \intT \Delta_T (x_j-x)J(x)-\lambda^2\sumj
\sum_{k<j} s_j s_k \Delta_T (x_j-x_k)\right\} .
\dle{sggenfun}
\ea
The above expression still contains both the UV ($\varepsilon\rightarrow
0^+$) and IR  ($\tilde{\mu}\rightarrow 0^+$) divergences of the
propagator. However, we will see that the observables and correlators
of interest are finite, after renormalisation,
 when those limits are taken and furthermore
that the $\tilde{\mu}\rightarrow 0^+$
limit selects a particular value of $k$ in (\dr{sggenfun}), which (see
below) is again equivalent to the fermion chiral selection rule.  
 
\subsubsection{The MT generating functional}

The MT model generating functional is defined, as customarily, as 

\be
Z^{MT}[\bar\eta,\eta;T,\mu]=N_\beta^F
\pif\exp\left[-\intT\left({\cal L}_{MT}[\bar\psi,\psi]+\mu j^0(x)
-\bar\eta\psi
-\bar\psi\eta\right)\right]
\label{mtgenfun1}
\ee
where, for completeness, 
 we have also  included a chemical potential $\mu$ term that we will analyse
 in section \dr{chempot}. 
 Following \dc{fugasc82,na85,zj,Belg}, let us  rewrite the quartic Thirring
interaction as a ``gauge-like'' interaction with an auxiliary vector
field $A_\mu$;
\be
Z^{MT}[\bar\eta,\eta;T,\mu] =\frac{N_\beta^F}{Z_A(T)}
\pivec \exp\left[-\frac{1}{2}\intT A_\mu (x)
A^\mu (x)\right] Z_f[A,\bar\eta,\eta;T,\mu]
\dle{auxpf}
\label{mtgenfun2}
\ee
with
\be
Z_f[A,\bar\eta,\eta;T,\mu]=
\pif    \exp\left[\intT\left(\bar\psi\left(\pabar+ig\Abar (x)+m_0
-\mu\gamma^0\right)\psi+\bar\eta\psi
+\bar\psi\eta\right)\right]
\dle{mtgenfun3}
\ee
and
$$
Z_A (T)=\pivec \exp\left[-\frac{1}{2}\intT A_\mu (x)
A^\mu (x)\right].
$$
In the remaining of this section and in section \ref{sec:piopequiv} we will 
 concentrate only in the case $\mu=0$. The $\mu\neq 0$ case will be studied 
 in section \dr{chempot}
\subsubsection{The partition functions}
%\dle{sec:piparfun}

We now turn to reproduce the result in \dc{Belg} for the partition
functions of the SG and MT models,  paying particular
attention to UV renormalisation and to the $\tilde{\mu}\rightarrow 0^+$
limit. 

Let us start with the SG partition function, which  is obtained by 
setting $J=0$ in (\dr{sggenfun}).  Using the explicit form of the
propagator in (\dr{bosprop}) one finds that $Z_{SG}$ is proportional to
$\tilde{\mu}^\alpha$ with 
$$
\alpha=\frac{\lambda^2}{4\pi}(\sumj s_j)^2=\frac{\lambda^2}{4\pi}(2k-n)^2.
$$
Hence, in the 
$\tilde{\mu}\rightarrow 0^+$ limit the only term that survives in the 
$k$-sum is $k=n/2$, which in turn selects only even values for $n$
in the $n$-sum.  Here we also see why, in the path-integral language,
the value of $K$ in (\dr{bosprop}) is irrelevant:  the reason is that 
any constant term in $\Delta_T$ cancels when $J=0$ in
(\dr{sggenfun}) since it is multiplied by $(2k-n)$ in the only surviving term
$k=n/2$.  Then, redefining variables as $x_j\rightarrow y_j$ for
$j=n/2+1,...,n$ and $n\rightarrow 2n$ and taking into account the
definition of the renormalised coupling constant $\alpha$ in
(\dr{renalpha}) we get
\ba
Z_{SG}(T)&=&Z_0^B(T)\sumn \left(\frac{1}{n!}\right)^2
\left[\frac{\alpha}{2\lambda^2}\left(\frac{T}
{\rho}\right)^{\lambda^2/4\pi}\right]^{2n}\nonumber\\
&\times&\prodjint 
\frac{\prod_{k<j}\left[
Q^2(x_j-x_k)Q^2(y_j-y_k)\right]^{\lambda^2/4\pi}}
{\prod_{k=1}^n\left[Q^2(x_j-y_k)\right]^{\lambda^2/4\pi}}
\dle{sgparfun2}
\ea
({\it c.f.} equation (\dr{sgparfun})). 
As in section \dr{ssubs:pt}, we again see that, at least for $\lambda^2<4\pi$, 
all the UV divergences of  the theory are removed by renormalising
$\alpha$ (see our comments in section \dr{ssubs:pt} about the regime 
 $4\pi\leq \lambda^2 < 8\pi$).

Our next step will be to calculate the MT  model partition
function and compare it with (\dr{sgparfun2}). First, we  
 write the auxiliary vector field $A_\mu (x)$ as 
\dc{na85,zj}
\be
A_\mu (x)= -\frac{1}{g}\left[ \partial_\mu\eta (x)+i\epsilon_{\mu\nu}
\partial_\nu \phi (x)\right]
\dle{vecfi}
\ee
with $\eta (x)$ and $\phi (x)$ two  boson fields, chosen to be 
periodic in Euclidean 
time. Then, by performing the following chiral transformation in the 
path integral fermion variables
\ba
\psi (x)&=&\exp\left\{ i\left[\eta(x)+\gamma_5 \phi(x)\right]
\right\}\psi' (x)\nonumber\\
\bar\psi (x)&=&\bar\psi' (x)
\exp\left\{ i\left[-\eta(x)+\gamma_5 \phi(x)\right]
\right\}
\nonumber
\ea 
the $\mu=0$ MT partition function, obtained by setting 
$\bar\eta=\eta=\mu=0$ in 
 (\dr{mtgenfun2})-(\dr{mtgenfun3}),  may be rewritten as
\ba
Z_{MT} (T)&=&\frac{N_\beta^F J_A}{Z_A(T)}
\pif\pib d\eta \exp\left\{\intT\left[ 
-\frac{1}{2g^2}\left[\partial_\mu\eta (x)\right]^2\right.\right.
\nonumber\\
&+&\left.\left.\frac{1}{2\kappa^2}\left[\partial_\mu\phi (x)\right]^2
+\bar\psi (x)\left[
\pabar+m_0 e^{2i\gamma_5\phi(x)}\right]\psi(x)\right]\right\}.
\dle{pfthprev}
\ea
Here $J_A$ is the Jacobian of the transformation $dA_\mu=J_A d\phi d\eta$, 
$\kappa$ is given by (\dr{kapdef}) and  we have used that 
the Jacobian of  the change of variable $d\bar\psi d\psi = J_F
d\bar\psi' d\psi'$,  which is given by the axial anomaly
\dc{na85,fuji84,zj}
$$
J_F=\exp\left[\frac{1}{2\pi}\intT \partial_\mu \phi (x)\partial^\mu \phi (x)
\right].
$$
Finally we have relabelled the fermion
variables $\psi '\rightarrow \psi$. 
Note that in (\dr{pfthprev}) the $\eta$ field is free and the only 
interaction term is the one proportional to $m_0$ which mixes 
the fermion fields and $\phi (x)$.  Now expand in powers
of $m_0$, note that $\bar\psi e^{2i\gamma_5 \phi}\psi=\sigma_+
e^{2i\phi}+\sigma_- e^{-2i\phi}$  and recall that, due to the fermion chiral
selection rule, only terms with the same number of
$\sigma_+$ and $\sigma_-$ contribute. 
The integral over $\eta$ is readily done and the boson and fermion 
correlators appearing in an arbitrary term in the expansion are of the
form (\dr{bosfreecorr}) and (\dr{ferfreem=0corr}), though when using
(\dr{bosfreecorr}) in (\dr{pfthprev}) the fields must be rescaled
$\phi\rightarrow i\kappa\phi$ because of the $(1/\kappa^2)$
factor in the kinetic term.  Finally, recalling that $Z_A(T)=J_A
(Z_0^B(T)/N_\beta)^2$ up to $T$-independent terms and then relabelling the
sum and integral variables gives
\ba
Z_{MT} (T)&=&Z_0^F (T)\sumn \left(\frac{1}{n!}\right)^2 
\left[\frac{m}{2\beta}\left(\frac{\rho}
{T}\right)^{\kappa^2/\pi}\right]^{2n}
\nonumber\\
&\times&\prodjint 
\frac{\prod_{k<j}\left[
Q^2(x_j-x_k)Q^2(y_j-y_k)\right]^{1-\kappa^2/\pi}}{
\prod_{k=1}^n\left[Q^2(x_j-y_k)\right]^{1-\kappa^2/\pi}}
\dle{thparfun}
\ea
where we have renormalised $m_0$ as  (\dr{renbarm}).  This is just equation
(\dr{thparfuna}) and shows how the work in \dc{Klaib} simplifies in 
the path integral framework.
Again by comparing (\dr{thparfun}) with (\dr{sgparfun2}) and using 
(\dr{pffree}) we reach the same conclusion as that of section \dr{ssubs:pft}
and \dc{Belg};
\be
Z_{MT}(T)=Z_{SG}(T)
\dle{pfequiv}
\ee 
up to vacuum energy and $T$-independent terms, provided the renormalised 
coupling constants of both models are identified as in
(\dr{equivconst1})-(\dr{equivconst2}).

At this point, notice that we could have expanded directly in $m_0$ 
in the original MT  Lagrangian, and then every term in the 
expansion would have been the $g\neq 0$, $m_0=0$ correlator of a product 
of $\sigpm$. Thus from (\dr{thparfun}) one obtains
\ba
\ll T_c \prodj \sigma_+ (x_j) \sigma_- (y_j) 
\gg_{\displaystyle\tiny\begin{array}{c}m=0\\g\neq 0\end{array}} 
&=&
 \left[\frac{( T\varepsilon)^{-\kappa^2/\pi}}{2\beta}\right]^{2n}
\nonumber\\&\times&
\prod_{j=1}^n
\frac{\prod_{k<j}\left[
Q^2(x_j-x_k)Q^2(y_j-y_k)\right]^{1-\kappa^2/\pi}}
{\prod_{k=1}^n\left[Q^2(x_j-y_k)\right]^{1-\kappa^2/\pi}}
\dle{ferm=0corr}
\ea
which, as we said in section \dr{renth}, is infinite.  However, with
the renormalisation (\dr{rensigma}) for the composite operators
$\sigpm$ one obtains (\dr{sigcorrel}) with  $M=\rho$ and so this is
just the $T>0$  extension of the result in \dc{Klaib}, as required (see
 our comments in section \dr{ren} about our renormalisation conventions).

Also once again comparison of (\dr{ferm=0corr}) with (\dr{bosfreecorr})
gives for the renormalised correlators
\be
\ll \prodj \sigma^R_+ (x_j) \sigma^R_- (y_j) 
\gg_{\displaystyle\tiny\begin{array}{c}m=0\\g\neq 0\end{array}}
=\left(\frac{\rho}{2}\right)^{2n}
\ll T_c 
\prodj A^R_+(x_j) A^R_- (y_j)\gg_0
%\exp\left[i\lambda\left(\phi(x_j)-\phi(y_j)\right)\right]
\dle{m=0nsigmaequiv}
\ee
as long as (\dr{equivconst1}) holds, which suggests the equivalence 
$\sigma_{\pm}=(\rho/2)\exp (\pm i\lambda\phi)$. This is the $m=\alpha=0$
version of (\dr{equivoper1}), just by combining it with 
(\dr{equivconst2}).  In the next section we will obtain the $m\neq 0$
version of this formula.

\section{The mass and current equivalences at $T>0$ and $\mu=0$}
\dle{sec:piopequiv}

In this section we show that the mass and current operator equivalences  
(\dr{equivoper1})-(\dr{equivoper2}) also  hold at finite temperature, 
i.e, for thermal averages of correlators of  those operators.  The 
condensates of $\sigma_\pm$ and $A_\pm$ are analysed in section
\dr{subsec:masscond} whereas we consider an arbitrary chiral invariant
chain  of those operators in section \dr{subsec:masschiral}.  The thermal
current equivalence (\dr{equivoper2}) is discussed in section
\dr{currequiv}, where we make use of point-splitting regularisation.

\subsection{The mass equivalence for the  condensates.}
\dle{subsec:masscond}

The simplest correlators to look at to prove (\dr{equivoper1}) would be 
$\ll  A_\pm (x)\gg_{SG}$ and \linebreak $\ll \sigpm \gg_{MT}$ which 
do not vanish for $m\neq 0$ because the chiral symmetry is
broken.  However, their relationship is actually a consequence of 
the partition functions
equivalence (\dr{pfequiv}). The reason is that they are independent of
$x$ simply because of the translation invariance of the theory so that
\ba
\frac{\alpha_0}{\lambda^2}\ll \left[\cos\lambda\phi (x)\right]_{bare}
\gg_{SG} \; =
\frac{\alpha}{\lambda^2}\ll \left[\cos\lambda\phi (x)\right]_R\gg_{SG}
\; =
\frac{1}{\beta L}\alpha\frac{\partial}{\partial\alpha} Z_{SG}(T)
\nonumber
\\
m_0 \ll \left[\bar\psi\psi\right]_{bare}\gg_{MT} \; =
m \ll \left[\bar\psi\psi\right]_R\gg_{MT} \; = \frac{1}{\beta L}m
\frac{\partial}{\partial m} Z_{MT} (T).
\nonumber
\ea
In fact, the above fermion correlator is nothing but the fermion
condensate which plays the r\^{o}le of the order parameter for the chiral
phase transition.  However, we do not expect chiral
symmetry restoration at finite temperature here since Coleman's
theorem prevents any continuous symmetry to be 
spontaneously broken in 2D \dc{col73}.
%\dnote{It would be nice if we 
%could see that directly in our expressions}. 

Therefore to prove the desired equivalence for the above correlators, all 
we have to do is to differentiate in (\dr{sgparfun2}) and
(\dr{thparfun}) with respect to $\alpha$ and $m$ respectively.  However,
it is instructive for the next sections to carry out an explicit
calculation.
%\dnote{CUT:?  and this, in
%turn,  provides us with a good check of consistency.}
From (\dr{thav}) and (\dr{sggf}),  we have
$$
\ll A_\pm \gg_{SG}\; =
\frac{Z_{SG}[J_\pm;T]}{Z_{SG}(T)} \quad \mbox{with} \quad 
J_\pm (x)=\mp i\lambda \delta^{(2)} (x-z).
$$
Next use (\dr{sggenfun}) to obtain the selection rule for this
correlator:  the total power of $\tilde{\mu}$ is now
given by $(\lambda^2/4\pi)(1\pm 2k-n)^2$, so that only 
$k=(n\mp 1)/2$ and hence  odd values of
$n$ in the sum, contribute now to (\dr{sggenfun}). Then, 
 by redefining $n\rightarrow 2n+1$, we obtain
\ba
\ll A_\pm (x)\gg_{SG}&=&
\frac{Z_0^B(T)}{Z_{SG}(T)}
\sumn \frac{1}{n!(n+1)!}
\left[\frac{\alpha}{2\lambda^2}\left(\frac{T}
{\rho}\right)^{\lambda^2/4\pi}\right]^{2n+1}
 \nonumber\\
&\times& ( T\varepsilon)^{\lambda^2/4\pi}\prod_{j=1}^{n}
\prod_{l=1}^{n+1}\int_T d^2 x_j d^2 y_l
\left[\frac{Q^2(x_j-x)}{Q^2(y_l-x)}\right]^{\lambda^2/4\pi}
\nonumber\\
 &\times&\frac{\prod_{k<j}\prod_{k'<l}\left[Q^2(x_j-x_k) Q^2(y_l-y_{k'})
\right]^{\lambda^2/4\pi}}{\left[Q^2(x_j-y_l)\right]^{\lambda^2/4\pi}}.
\dle{expequivprev}
\ea
Here we have relabelled $x_j\longleftrightarrow y_l$ and used 
$Q^2(x)=Q^2(-x)$. 

Observe that $\ll A_\pm (x) \gg$ is the same for both $\pm$, 
which we could have expected by realising that both the SG Lagrangian and
the path integral measure  are invariant under $\phi\rightarrow -\phi$.
Also the UV $\varepsilon$ dependence in (\dr{expequivprev}) can be 
absorbed in the renormalisation of $\exp (i\lambda\phi)$
according to (\dr{renexp}) so that once again 
there are no more infinities for $\lambda^2<4\pi$ (see comments above). 
Thus we are allowed to shift the integration variables as
$x_j\rightarrow x_j +x$ and $y_j\rightarrow y_j +x$ $\forall j$, which 
shows explicitly that the above correlator is actually $x$ independent.
The final answer is
%\dnote{We were talking in Madrid about removing one 
%of (\dr{expequivprev}) and (\dr{sgexp}). I haven't done it but I could 
% sacrifice one if the paper gets too long}
%
%
\ba
\ll A^R_\pm (x)\gg_{SG}&=& 
\frac{Z_0^B(T)}{Z_{SG}(T)}
\sumn \frac{1}{n!(n+1)!}
\left(\frac{\alpha}{2\lambda^2}\right)^{2n+1}
\nonumber\\
&\times& 
\left(\frac{T}
{\rho}\right)^{(n+1)\lambda^2/2\pi}
\prod_{j=1}^{n}
\prod_{l=1}^{n+1}\int_T d^2 x_j d^2 y_l
\left[\frac{Q^2(x_j)}{Q^2(y_l)}\right]^{\lambda^2/4\pi}
\nonumber\\
 &\times&\frac{\prod_{k<j}\prod_{k'<l}\left[Q^2(x_j-x_k) Q^2(y_l-y_{k'})
\right]^{\lambda^2/4\pi}}{\left[Q^2(x_j-y_l)\right]^{\lambda^2/4\pi}}
\dle{sgexp}
\ea
which vanishes in the free case $\alpha=0$, 
consistently with the discussion in sections \dr{ssubs:pt} and 
\dr{subs:pigf}. It can also be checked, order by order in the $\alpha$ 
expansion that the above formula indeed gives the same answer as 
 $(\lambda^2/\beta L)(\partial Z_{SG}/\partial\alpha)$,
 as explained above, which is a consistency check. 
%\dnote{Actually, I checked 
% it only for the first order. We can remove this comment if 
% you wish. As it is written now I am not lying though. } 
 
Now calculate the  $\sigpm$ correlator in the massive Thirring 
model. Following the same steps as those leading from  (\dr{auxpf}) to 
(\dr{pfthprev}) and taking into account that 
 $$\bar\psi\exp[2i\gamma_5
\phi]P_\pm\psi=\sigpm\exp[\pm 2i\phi]$$
 gives
\ba
\ll \sigpm \gg_{MT}&=&\frac{Z_0^F(T)}{Z_{MT}(T)}\sumn\frac{m_0^n}{n!}
 \sum_{k=0}^n \nk 
\left[\prod_{j=1}^{k}\prod_{l=1}^{n-k}\int_T d^2 x_j d^2 y_l\right]
\nonumber\\
&\times& \ll T_c \sigpm\prod_{j=1}^{k}\prod_{l=1}^{n-k}
\sigma_+ (x_j)\sigma_- (y_l)\gg_0 
\nonumber\\
&\times&
\ll T_c  e^{\mp 2\kappa\phi (x)}\prod_{j=1}^{k}\prod_{l=1}^{n-k} 
e^{-2\kappa\left[\phi(x_j)-\phi(y_l)\right]}\gg_0 .
\nonumber
\ea
The chiral selection rule then forces the choice $k=(n\mp 1)/2$ and 
$n\ge 1$  odd so as to  have the same number of $\sigma_+$ and $\sigma_-$. 
Now let $n\rightarrow 2n+1$, use  (\dr{bosfreecorr}) and 
(\dr{ferfreem=0corr}) and renormalise the mass and the composite 
operator $\sigpm$ as in (\dr{rensigma})-(\dr{renbarm}).  One obtains
\ba
\ll \sigpmr \gg_{MT}&=&
\frac{Z_0^F(T)}{Z_{MT}(T)}\sumn \frac{1}{n!(n+1)!}
\frac{\rho}{2}\left(\frac{m\rho}{2}\right)^{2n+1}
\left(\rho\beta\right)^{-2(1-\kappa^2/\pi)(n+1)}
\nonumber\\
 &\times&\prod_{j=1}^{n}
\prod_{l=1}^{n+1}\int_T d^2 x_j d^2 y_l
\left[\frac{Q^2(x_j)}{Q^2(y_l)}\right]^{1-\kappa^2/\pi}
\nonumber\\
 &\times&\frac{\prod_{k<j}\prod_{k'<l}\left[Q^2(x_j-x_k) Q^2(y_l-y_{k'})
\right]^{1-\kappa^2/\pi}}{\left[Q^2(x_j-y_l)\right]^{1-\kappa^2/\pi}}
\dle{thsigpm}.
\ea
Comparison of (\dr{thsigpm}) with (\dr{sgexp}) and use of
the equivalences (\dr{equivconst1}), (\dr{equivconst2}),
(\dr{pffree}) and  (\dr{pfequiv}) finally gives
\be
\ll \sigpmr \gg_{MT} \; =\frac{\rho}{2}
\ll A^R_\pm (x) \gg_{SG}.
\dle{sigmaexpequiv}
\ee
As this identity depends explicitly on the scale $\rho$, it can be
combined with (\dr{equivconst2}) to get
\be
m\ll\left[\bar\psi(x)\psi(x)\right]_R\gg_{MT} \; =\frac{\alpha}{\lambda^2}
\ll\left[\cos\lambda\phi(x)\right]_R\gg_{SG}
\dle{cosequiv}
\ee
which is scale independent.

\subsection{The mass equivalence for chiral invariant operators}
\dle{subsec:masschiral}

We now prove  (\dr{equivoper1}) for thermal
correlators of an arbitrary number of $\sigma_\pm$ operators evaluated at
different space-time points. As was commented previously, there is no need 
to consider the same number of $\sigma_+$ and $\sigma_-$ in the
$m\neq 0$ case. However, such a chiral
invariant combination is chosen here because then the selection 
rule will be just the
same as for $m_0=0$ which clearly simplifies matters.  We do not
present the calculation for more complicated
cases here since it is more involved but it goes through in the same way and
leads to the expected equivalence, as we will comment later.

Once again we start from the SG
model and calculate the correlator
$$
A_N (X;Y)\equiv
\ll T_c \prodjN   A_+(X_j) A_- (Y_j)\gg_{SG} \; =
%\exp\left[i\lambda\left(\phi(X_j)-\phi(Y_j)\right)\right]
\frac{Z_{SG}[J_N;T]}{Z_{SG}(T)}
$$
with
\be
J_N (z)=-i\lambda\sum_{j=1}^N\left[\delta^{(2)} (z-X_j)-
\delta^{(2)} (z-Y_j)\right].
\dle{jn}
\ee
%\dnote{Cut:
%Notice that the above correlators are the only ones that are evaluated
%in \dc{Coleman} and in section \dr{ssubs:pt}.}  
Notice that 
differentation of the partition
function would only give global information about  the
above correlator, i.e, its total integral over  $X_j$ and $Y_j$, which is
not enough to prove the required equivalence. 
As before, on substitution of  (\dr{jn}) into (\dr{sggenfun}) only 
$k=n/2$ survives. On the other hand, $Z_0^B[J_N,T]$ gives precisely 
the free contribution in (\dr{bosfreecorr}) so that
\ba
A_N^R (X;Y)
&=&\frac{Z_0^B (T)}{Z_{SG} (T)}
 A_N^R(X;Y)_{0}
\sumn \left(\frac{1}{n!}\right)^2
\left[\frac{\alpha}{2\lambda^2}\left(\frac{T}
{\rho}\right)^{\lambda^2/4\pi}\right]^{2n}
\nonumber
\\
&\times&\prodjint 
\frac{\prod_{k<j}\left[
Q^2(x_j-x_k)Q^2(y_j-y_k)\right]^{\lambda^2/4\pi}}
{\prod_{k=1}^n\left[Q^2(x_j-y_k)\right]^{\lambda^2/4\pi}}
\nonumber
\\
&\times&
\prod_{k=1}^N\left[\frac{Q^2(x_j-X_k)Q^2(y_j-Y_k)}
{Q^2(x_j-Y_k)Q^2(y_j-X_k)}\right]^{\lambda^2/4\pi}
\nonumber
\ea
where, according to (\dr{bosfreecorr}) and (\dr{renexp}),
\be
 A^R_N(X;Y)_{0}=
\left(\frac{T}{\rho}\right)^{N\lambda^2/2\pi}\prod_{j=1}^N
\frac{\prod_{k<j}\left[
Q^2(X_j-X_k)Q^2(Y_j-Y_k)\right]^{\lambda^2/4\pi}}
{\prod_{k=1}^N\left[Q^2(X_j-Y_k)\right]^{\lambda^2/4\pi}}.
\dle{0exps}
\ee
Now, in the Thirring model we have
%\dnote{The 3 first lines of this
%formula could be removed, if necessary}
%
\ba
\lefteqn{ \ll T_c  \prodjN\sigma^R_+ (X_j)\sigma^R_-(Y_j)\gg_{MT}}
\nonumber
\\
&=&
(\varepsilon\rho)^{2N\kappa^2/\pi}\frac{Z_0^F (T)}{Z_{MT} (T)}
 \sumn \frac{m_0^{2n}}{(n!)^2}
\left[\prod_{j=1}^{n+N}\int_T d^2 x_j d^2 y_j \right]
\left[\prod_{k=1}^{N}\delta^{(2)}(x_k-X_k)\delta^{(2)}(y_k-Y_k)
\right]
\nonumber
\\
&\times&
\ll T_c \prod_{j=1}^{n+N} 
\sigma_+ (x_j)\sigma_- (y_j)\gg_0\ll T_c \prod_{j=1}^{n+N} 
e^{-2\kappa\left[\phi (x_j)
 -\phi (y_j)\right]}\gg_0
\nonumber
\\
&=&
\frac{Z_0^F (T)}{Z_{MT} (T)}
\ll T_c \prodjN\sigma^R_+ (X_j)\sigma^R_-(Y_j)
\gg_{\displaystyle\tiny\begin{array}{c}m=0\\g\neq 0\end{array}}
\sumn \left(\frac{1}{n!}\right)^2 
\left[\frac{m}{2\beta}\left(\frac{\rho}
{T}\right)^{\kappa^2/\pi}\right]^{2n}
\nonumber
\\
&\times&
\prodjint 
\frac{\prod_{k<j}\left[
Q^2(x_j-x_k)Q^2(y_j-y_k)\right]^{1-\kappa^2/\pi}}{
\prod_{k=1}^n\left[Q^2(x_j-y_k)\right]^{1-\kappa^2/\pi}}
\nonumber\\
&\times &
\prod_{k=1}^N\left[\frac{Q^2(x_j-X_k)Q^2(y_j-Y_k)}
{Q^2(x_j-Y_k)Q^2(y_j-X_k)}\right]^{1-\kappa^2/\pi}
\ea
where, from (\dr{ferm=0corr}) and (\dr{rensigma})
\ba
\ll T_c \prodjN\sigma^R_+ (X_j)\sigma^R_-(Y_j)
\gg_{\displaystyle\tiny\begin{array}{c}m=0\\g\neq 0\end{array}}=
\left[\frac{1}{2\beta}
\left(\frac{\rho}{T}\right)^{\kappa^2/\pi}\right]^{2N}
\nonumber\\\times
\prod_{j=1}^N\frac{\prod_{k<j}\left[
Q^2(X_j-X_k)Q^2(Y_j-Y_k)\right]^{1-\kappa^2/\pi}}
{\prod_{k=1}^N\left[Q^2(X_j-X_k)\right]^{1-\kappa^2/\pi}}
\dle{m=0sigmas}.
\ea
Therefore, using (\dr{m=0nsigmaequiv}) and the equivalences between the
coupling constants and partition functions, we find the desired result
$$
\ll T_c \prodjN\sigma^R_+ (X_j)\sigma^R_-(Y_j)\gg_{MT} \;=
\left(\frac{\rho}{2}\right)^{2N}
\ll T_c \prodjN   A^R_+(X_j) A^R_- (Y_j)\gg_{SG}.
$$
For more complicated strings of $\sigma_{\pm}$ the
calculation goes through in a similar way and leads to the expected
result in which $\sigma^R_{\pm} (x_j) \rightarrow (\rho/2)
A_{\pm}^R(x_j)$.

\subsection{The current equivalence}
\dle{currequiv}

We now analyse the current equivalence (\dr{equivoper2}). 
When dealing with correlators including insertions of the current operator 
$\curr$ in the Thirring model, there is an added complication since 
a product of fields at the 
same point gives a divergence.  We thus have to specify how  to 
regularise and, eventually, renormalise.  To do that we 
follow the same approach as in \dc{AlvGom98,rualv87} for the massless case
and
use point-splitting regularisation. Indeed, as we will
 expand in the fermion mass, 
it will be enough to regularise the current  for $m=0$.  Other
than that, the previous discussion about the renormalisation of the
theory will remain unaltered, i.e, no further  infinities will appear once 
 we choose the proper regulator (see below).  
We begin by quickly sketching how the point-splitting method is
applied to this case.

\subsubsection{Point-Splitting regularisation for the fermion current}

The point-splitting regularisation method \dc{fried72} consists in
replacing  products of fermion fields $\bar\psi_\alpha (x)\psi_\beta
(x)\rightarrow \lims \bar\psi_\alpha (y)\psi_\beta (x)$ where 
 $\lims f(x,y)\equiv (1/2)\lim_{y\rightarrow
x}[f(x,y)+f(y,x)]$ is the symmetric limit. 
Clearly we can multiply any fermion bilinear by any 
function $F(x,y)$ such that $F(x,x)=1$. As usual, 
$F(x,y)$ can be  determined by
demanding that the different observables obey physical conditions such
as current conservation or gauge
invariance  \dc{fried72}.  Consider for example 
the current $\curr$, which is the relevant one here, and suppose we
 just choose  $F(x,y)=1$. 
 Recall that the relevant theory here, according to 
(\dr{mtgenfun3}) for $\mu=0$, is 
that of a fermion field in the presence of an external vector field 
$A_\mu (x)$.
Then classically, $\curr$ is a conserved current in such a theory.
However, it is easy to check that
   $\partial^\mu_x \lims \bar\psi (y)\gamma_\mu\psi(x)=
\lim_{y\rightarrow x} (1/2) A_\mu(x)[\bar\psi (y)\gamma_\mu\psi (x)-
\bar\psi (x)\gamma_\mu\psi (y)]$, i.e, that choice for $F(x,y)$ would not lead 
 to a classically conserved current. 
  We see instead that the correct 
regularised  current is
\be
\curreg={\lim_{y\rightarrow x}}^s 
\bar\psi (y)\gamma_\mu \psi (x)\exp\left[ig\int_{y}^{x}
 A_\alpha(\xi)d\xi^\alpha\right]
\dle{curreg}
\ee
which is classically conserved before taking the symmetric  limit. 
Also (\dr{curreg}) can be seen to satisfy the Ward identity 
$\partial^\mu\ll \curreg \gg=0$, just by writing
$\ll\bar\psi_\alpha(x)\psi_\beta (y)\gg=G^>_{\alpha\beta} (x,y)$, 
the advanced  exact fermion propagator, and, in addition,  
% with
%$G^>_{\alpha\beta} (x,y)=\theta (x^0-y^0)G_{\alpha\beta} (x,y)$ and
%$G_{\alpha\beta}$ the exact 
%fermion propagator satisfying $[\pabar_x+ig
%\Abar (x)]G(x,y)=\delta^{(2)} (x-y)$
 when dealing with a gauge theory, this definition
of the regularised current is gauge invariant,  another requirement
the current should satisfy \cite{AlvGom98,rualv87,fried72}. 

We note that in the free case $g=0$, the point-split current yields 
automatically a finite result for correlators with current insertions,
without any need of renormalisation. The  reason is that, from
(\dr{freeferprop12}), we see that the divergent part of $S_\pm$ 
 always behaves like $(\epsilon_0\pm\epsilon_1)^{-1}$
with $\epsilon_\mu=y_\mu-x_\mu$ and hence it cancels when the
symmetric limit is taken. Furthermore, we will now see that those are the
only infinities also for $g\neq 0$ so that no extra
renormalisations are needed and the choice (\dr{curreg}) for $F(x,y)$ is the
only one giving  the right equivalence with the SG model.

\subsubsection{Chiral invariant mixed correlators}

The Thirring model Lagrangian and the fermion measure are invariant under
charge conjugation $C$ and therefore $\ll j_\mu (x) \gg_{MT} \; =0$, 
consistently
with \mbox{$\epsilon_{\mu\nu}\partial_\nu\ll\phi(x)\gg_{SG}=0$}, 
following just from the $\phi\rightarrow -\phi$ invariance of the SG
model. Thus, the equivalence is too simple with just one insertion of the
current. In this section we concentrate on the mixed propagator with one
current and one $\sigma_+\sigma_-$ insertion.  This correlator again has
the useful property that it is chiral invariant so that the selection
rules are very simple.  Notice that our results can then be extended
without difficulty to an arbitrary string of $ \sigma_+\sigma_-$ and/or 
 $j_\mu$'s or even
to chiral non-invariant combinations,  but for clarity we restrict here to the
simplest example.  In turn, we will be showing  that the
equivalence is also true for mixed mass-current correlators. 

Thus, according to the discussion in the previous section, consider
 the correlator 
$$
F_\mu (x,z_1,z_2)=\ll T_c  \curreg \sigma_+ (z_1) \sigma_- (z_2)\gg_{MT}
$$
with $\curreg$ in (\dr{curreg}). After similar manipulations as those 
of the previous sections one obtains
\ba
F_\mu (x,z_1,z_2)&=&\frac{Z_0^F (T)}{Z_{MT} (T)}
 \lims \sumn \frac{m_0^{2n}}{(n!)^2}
\left[\prod_{j=1}^{n+1}\int_T d^2 x_j d^2 y_j \right]
\delta^{(2)}(x_1-z_1)\delta^{(2)}(y_1-z_2)
\nonumber\\
&\times&
 \sum_{\pm}
 \ll T_c \bar\psi (y)\gamma_\mu P_\pm \psi (x)\prod_{j=1}^{n+1}
\sigma_+ (x_j)\sigma_- (y_j)\gg_0\nonumber\\
&\times&
\ll T_c e^{\mp\kappa [\phi(x)-\phi(y)]}
e^{i\kappa\epsilon_{\alpha\beta}\int_{y}^{x} 
\partial^\beta \phi(\xi)d\xi^\alpha}\prod_{j=1}^{n+1}
e^{-2\kappa\left[\phi (x_j) -\phi (y_j)\right]}\gg_0
\dle{curreq1}
\ea
where  the $\eta$ field dependence is the same as  in (\dr{pfthprev})
and has been integrated out.  Now notice that the only pole in 
$(x-y)_\mu \equiv \epsilon_\mu\rightarrow 0$ is in the fermion 
correlator in (\dr{curreq1}) and that it goes like $\epsilon^{-1}$. 
Thus the first two exponentials in the boson free 
correlator in the last line can be expanded 
up to $\Od (\epsilon)$.  Let us analyse first 
the fermion free thermal correlator in  (\dr{curreq1}).  After shifting
$\bar\psi\rightarrow\bar\psi\gamma^0$, all the required fermion 
correlators are of the form
\ba
(-1)^{n+1}\ll T_c 
\left[ \bar\psi_1 (x)\psi_1 (y)\pm  \bar\psi_2 (x)\psi_2 (y)\right]
\prod_{j=1}^{n+1}\bar\psi_1 (y_j)\psi_1 (x_j) \bar\psi_2 (x_j) 
\psi_2 (y_j) \gg_0\nonumber\\
 =\left(-\frac{1}{2\beta}\right)^{2n+3}\frac{1}{Z^+_Q(\epsilon)}
\prod_{j=1}^{n+1}\frac{\prod_{k<j}\left[
Q^2(x_j-x_k)Q^2(y_j-y_k)\right]}
{\prod_{k=1}^{n+1}\left[Q^2(x_j-y_k)\right]}
\nonumber\\
\times \left[1+\epsilon^\mu\partial_\mu\log\prod_{l=1}^{n+1}
\frac{Z^+_Q (x_l-x)}{Z_Q^+(y_l-x)}\right] \pm 
\left(\begin{array}{c}x_j\longleftrightarrow y_j\\
Z_Q^+\longleftrightarrow Z_Q^-\end{array}\right)+\Od (\epsilon^2)
\dle{curreq2}
\ea
where $Z_Q^\pm$ are defined in (\dr{zq}), we have used (\dr{wickfer})
and (\dr{det}) and we have expanded to the relevant order in $\epsilon$. 
As for the boson correlators, to $\Od (\epsilon)$, we only need to
calculate 
\ba
\ll T_c \partial_\mu^x \phi (x)\prod_{j=1}^{n+1}\exp 
\left[-2\kappa( \phi (x_j)-\phi (y_j))\right]\gg_0=-\frac{1}{Z_0^B(T)}
  \partial_\mu^x \left.\frac{\delta}{\delta J} 
Z_0[J;T]\right\vert_{J=J_\kappa}
\nonumber\\
=\frac{(\varepsilon T)^{-2(n+1)\kappa^2/\pi}}{2\pi}
\prod_{j=1}^{n+1}\frac{\prod_{k<j}\left[
Q^2(x_j-x_k)Q^2(y_j-y_k)\right]^{-\kappa^2/\pi}}
{\prod_{k=1}^{n+1}\left[Q^2(x_j-y_k)\right]^{-\kappa^2/\pi}}
\partial_\mu^x\log\prod_{l=1}^{n+1}\frac{Q^2 (x-x_l)}{Q^2(x-y_l)}
\nonumber
\ea
where
$$
 J_\kappa(z)
=-2i\sum_{j=1}^{n+1}\left[\delta^{(2)}(z-x_j)-\delta^{(2)}(z-y_j)\right].
$$
Notice that by taking the partial derivative $\partial^x_\mu$ we make sure 
that there is no extra $\tilde{\mu}$-dependence coming from the
propagator.  Also recall that the UV $\varepsilon$ factor above (not to
be confused with $\epsilon_\mu$) will be absorbed in the
renormalisation of the mass and the $\sigma_\pm$ fields as indicated in
previous sections. Finally, use of
(\dr{propQ})  in (\dr{curreq2}) gives
\ba
F^R_\mu (x,z_1,z_2)
=
\frac{i}{2\pi}\left(1-\frac{\kappa^2}{\pi}\right)\epsilon_{\mu\nu}
\frac{Z_0^F (T)}{Z_{MT} (T)}\sumn \left(\frac{1}{n!}\right)^2 
\left[\frac{m}{2\beta}\left(\frac{\rho}
{T}\right)^{\kappa^2/\pi}\right]^{2n}
\nonumber\\
\times
\ll T_c \sigma^R_+ (z_1)\sigma^R_-(z_2)
\gg_{\displaystyle\tiny\begin{array}{c}m=0\\g\neq 0\end{array}}
 \prodjint
\frac{\prod_{k<j}\left[
Q^2(x_j-x_k)Q^2(y_j-y_k)\right]^{1-\kappa^2/\pi}}{
\prod_{k=1}^n\left[Q^2(x_j-y_k)\right]^{1-\kappa^2/\pi}}
\nonumber\\
\times
\left[\frac{Q^2(x_j-z_1)Q^2(y_j-z_2)}
{Q^2(x_j-z_2)Q^2(y_j-z_1)}\right]^{1-\frac{\kappa^2}{\pi}}\!\!\!\!
\partial_x^\nu\log \frac{Q^2 (x-z_1)}{Q^2(x-z_2)}
\prod_{l=1}^{n}\frac{Q^2(x-x_l)}{Q^2(x-y_l)}
\dle{curreqth}
\ea
with $\ll T_c \sigma^R_+ (z_1)\sigma^R_-(z_2)
\gg_{\displaystyle\tiny\begin{array}{c}m=0\\g\neq 0\end{array}}$ 
in (\dr{m=0sigmas})
for $N=1$. Notice that, thanks to (\dr{propQ}), the above result is
independent of the way we take the limits $\epsilon_0\rightarrow 0$ and
$\epsilon_1\rightarrow 0$, which is  a good check of consistency.

On the other hand, in the SG model we have to evaluate 
\ba
\ll T_c  \partial_\mu^x\phi (x) A^R_+ (z_1) A^R_- (z_2)\gg_{SG}
=-(\varepsilon\rho)^{-\lambda^2/2\pi}
\left.\frac{\delta}{\delta J}Z_{SG}[J;T]
\right\vert_{J=J_1}
\nonumber\\
=\frac{i\lambda}{4\pi}\frac{Z_0^B(T)}{Z_{SG}(T)}A^R_1 (z_1,z_2)_{0}
\sumn \left(\frac{1}{n!}\right)^2
\left[\frac{\alpha}{2\lambda^2}\left(\frac{T}
{\rho}\right)^{\lambda^2/4\pi}\right]^{2n}\nonumber\\
\times
\partial_\mu^x
\prodjint 
\frac{\prod_{k<j}\left[
Q^2(x_j-x_k)Q^2(y_j-y_k)\right]^{\lambda^2/4\pi}}
{\prod_{k=1}^n\left[Q^2(x_j-y_k)\right]^{\lambda^2/4\pi}}
\nonumber\\
\times 
\left[\frac{Q^2(x_j-z_1)Q^2(y_j-z_2)}
{Q^2(x_j-z_2)Q^2(y_j-z_1)}\right]^{\lambda^2/4\pi}\partial_\mu^x
\log\frac{Q^2 (x-z_1)}{Q^2(x-z_2)}\prod_{l=1}^{n}\frac{Q^2(x-x_l)}{Q^2(x-y_l)}
\dle{curreqsg}
\ea
with $J_1(z)=-i\lambda(\delta^{(2)}(z-z_1)-\delta^{(2)}(z-z_2))$ and 
$A^R_1 (z_1,z_2)_{0}$ as in (\dr{0exps}). Thus comparison of
(\dr{curreqsg}) with (\dr{curreqth}) and use of the equivalences we
have already proved gives
$$
\ll T_c \curreg \sigma^R_+ (z_1) \sigma^R_- (z_2)\gg_{MT}
=\left(\frac{\rho}{2}\right)^2\frac{\lambda}{2\pi}\epsilon_{\mu\nu}
\ll T_c \partial_\nu^x\phi (x)
 A^R_+ (z_1) A^R_- (z_2)\gg_{SG}
$$
which proves the current equivalence for this correlator. 

\section{Bosonisation of the massive Thirring model at\\ nonzero
chemical potential}\dle{chempot}

In this section we study the thermal bosonisation of the Thirring 
 model at nonzero chemical potential $\mu$. 
 The chemical potential is the Lagrange multiplier associated with the 
conservation of the number of fermions minus anti-fermions 
 (net fermion 
number) $Q_F=\int dx j^0(t,x)$. Hence, the Lagrangian density of the 
$\mu\neq 0$ Thirring model is just 
\be   
 {\cal{L}}_{MT} [\bar\psi,\psi;\mu] =  -\bar{\psi} (\pabar + m_0)\psi +
 \frac{1}{2}g^2  j_\alpha (x) j^\alpha (x)+\mu j^0 (x) , 
\dle{thchempot}
 \ee    
By virtue of the current
equivalence we have proved in section \dr{currequiv}, we can say that at
$\mu=0$, $\ll Q_F \bullet\gg_{MT} \; =\ll Q_K \bullet\gg_{SG}$, where 
$$
Q_K=\intsp \frac{\lambda}{2\pi}\frac{\partial}{\partial x^1}\phi (x)
=\frac{\lambda}{2\pi}\left[\phi(t,+\infty)-\phi(t,-\infty)\right]
$$
corresponds classically to the conserved number of kinks minus anti-kinks
\dc{colebook}.  Recall that classically $\phi_{cl}(t,\pm\infty)=2\pi
n_{\pm}/\lambda$, with $n_{\pm}\in\IZ$, so that $Q_K
[\phi_{cl}]=n_+-n_-$.  We regard this equivalence  as the thermal
identification between Thirring fermions and SG kinks at $T\neq 0$, thus
extending the $T=0$ standard result \dc{Coleman,colebook}, discussed 
in the introduction. Physically, we interpret it  as the existence of
fermion excitations (the kinks) in the thermal bath. 

The presence of excitations with net fermion number becomes crucial when 
chemical potentials are included and hence the grand-canonical 
ensemble partition function is considered. 
As the averaged net fermion number density
$\rho(\mu)=(\beta L)^{-1}\ll Q_F\gg$ will in general be different from
zero, a natural question to ask within this context is  what is the
bosonised version of the theory for $\mu\neq 0$. As we commented in the
introduction, the answer has been given in \dc{AlvGom98} for the
massless case, where the free boson partition function acquires an extra 
$\mu$-dependent term,
%\dnote{CUT: We recall that for $\alpha=0$ the classical 
%kinks are massless.},
unlike, for instance,  the massless Schwinger model, 
where there are no elementary fermion excitations in the bosonised theory 
(all the electric charge is confined), 
even for nonzero $T$ and $\mu$ \dc{AlvGom98}. The Schwinger model 
 partition function is then  equivalent to
that of a boson field with mass $e/\sqrt{\pi}$, with $e$ the electric
charge,  and  the chemical potential dependence vanishes in the  
partition function, yielding  $\rho (\mu)=0$. 

Our objective in this section is then  to analyse the partition
function of the $\mu\neq 0$  massive Thirring model (\dr{thchempot}), 
\be
Z_{MT}(T,\mu)=N_\beta^F\pif \exp \left[-\intT 
{\cal{L}}_{MT} [\bar\psi,\psi;\mu]\right].
\dle{zthmu}
\ee
The simplest way to proceed is to make use of the current equivalence
proved in section \dr{currequiv}.  Formally the argument goes as follows: we
expand $Z_{MT}(T,\mu)$ in $\mu$, so that  every term in the
expansion is a $\mu=0$ correlator of a product of $j^0(x_j)$ at different
space-time points $x_j$. Then assume that the $T\neq 0$ current
equivalence (\dr{equivoper2}) holds for any  arbitrary product of
currents so that we conclude  
\be
Z_{MT}(T,\mu)=Z_{SG\mu}(T,\mu)
\dle{parfuneqchempot}
\ee
with
\be
Z_{SG\mu}(T,\mu)=N_\beta \pib \exp\left[-\intT \left({\cal L}_{SG}
 -\frac{\mu\lambda}{2\pi}\frac{\partial}{\partial x^1}
 \phi(x)\right)\right]\quad ,
\dle{zsgmu}
\ee
i.e, the bosonised action is the SG model plus an extra term which is 
topological, in the sense that it only depends on the value of the field 
at the spatial boundary ($x=\pm\infty$).  As we have commented before, 
that term is interpreted as the result  of excitations with net kink
(fermion) charge being present in the thermal  bath and having
associated   the  chemical potential $\mu$ in the  grand-canonical
ensemble.  Recall that  an  analogous contribution was found 
in the chiral Lagrangian for low-energy QCD \dc{AlvGom95} where the
r\^{o}le of kinks (Thirring fermions)  is played by the skyrmions (QCD
baryons):  the chiral Lagrangian at finite baryon density acquires, among
other things, a new factor $\mu Q_{SK}$ 
with $Q_{SK}$ the skyrmion topological 
charge, analogously to (\dr{zsgmu}). The existence of this term provides
 an alternative proof, based on Statistical Mechanics, of the old result
 that the classical kinks have their quantum counterpart in a particle
 with fermion number equal to one \cite{Coleman,goja75} and it is 
 indeed another example of $\mu$-dependent topological terms found in the 
 literature \dc{rewi85}.

The equivalence (\dr{parfuneqchempot})-(\dr{zsgmu}) is the central
result of this section. We want to stress that, formally, 
 we have proved it only 
in perturbation theory around the $m=\alpha=0$ theory. 
Indeed, we can perform a simple 
consistency check of
(\dr{parfuneqchempot}), since   for $\alpha=0$, $Z_{SG\mu}$ should give the
 answer for the massless Thirring model obtained in
\dc{AlvGom98}. In that case,  ${\cal L}_{SG}$ in (\dr{zsgmu}) is just a
free boson Lagrangian, so let
\be
\phi (x,t)\rightarrow \phi (x,t) +\frac{\mu\lambda}{2\pi}x.
\dle{mushift}
\ee
As for the field measure, in the free theory the field is allowed to have 
$\Delta\phi\equiv \phi(L/2,t)-\phi(-L/2,t)=c$ with $c\in\IR$ arbitrary
and $L$ the length of the system, so that  the total energy still 
remains finite when $L\rightarrow\infty$.  In other words, the free
theory has trivial topology since any vacuum configuration can be reached
by infinitesimal variations of the field. Therefore, we can write the
path integral measure formally as
$\int_{\IR} dc\int_{\Delta\phi=c}d\phi$. Thus, under  (\dr{mushift}), 
$\Delta\phi\rightarrow \Delta\phi + \mu\lambda L/2\pi$, but then, by 
shifting $c\rightarrow c-\mu\lambda L/2\pi$, the path integral measure
remains invariant. Therefore, by completing the square and using 
(\dr{equivconst1}), we find
\be
Z_{SG\mu}^{\alpha=0}(T,\mu)=\exp\left[\beta L\frac{\mu^2}{2(\pi + g^2)}
\right] Z^B_0 (T)
\dle{zthalvgom}
\ee
which is exactly the result found in \dc{AlvGom98} for the massless 
Thirring model partition function.

In what follows, we provide an alternative 
way of obtaining the result (\dr{zsgmu}) 
without having to expand in $\mu$. For that purpose, we will make use 
 of some of the results obtained in \cite{AlvGom98} for the massless case. 
 Hence, let us  return to the massive theory for $\mu\neq 0$ and  
compare the series expansion of (\dr{zthmu}) in $m$  with that of
(\dr{zsgmu}) in $\alpha$, in the same way as we did in previous sections 
for $\mu=0$. Notice that we expand now around the massless $\mu\neq 0$
theory, without expanding in $\mu$. An important remark is in order
here: for every term in the $\alpha$-expansion, which  is an $\alpha=0$ 
boson  correlator, we will consider  that the measure is translation
invariant, as  discussed before.  In the mass
expansion of the $\mu\neq 0$ Thirring model, we will make the same
assumption whenever a boson correlator appears.  Therefore, it should
become clear that we are perturbing  around both the
$\alpha=0$ Lagrangian and path-integral measure and hence our results 
 are valid only in perturbation theory. 
%\dnote{CUT: This seems to be consistent with the 
% fact that for $\lambda\gg 1$ the spacing between the 
%classical potential zeroes becomes very small, so that it would be 
% justified to replace a sum over kinks by an integral.}
%\dnote{CUT:  This also corresponds 
%to a high density regime $\mu\gg\sqrt{\alpha}$ in which  the kinks overlap.} 
It remains unclear to us whether we would arrive to the same conclusion, 
 namely (\dr{zsgmu}), 
within a semi-classical approximation, i.e, perturbing now  around a
nontrivial classical configuration, which would be consistent for  
$\lambda\ll 1$ \dc{Coleman,goja75} but that is beyond the scope of this work
 and it will be analysed elsewhere.

Therefore, as in previous sections, let us introduce the auxiliary field 
 $A_\alpha$ and expand in $m_0$ in (\dr{zthmu}).  Then we have
\ba
Z_{MT} (T,\mu)&=&\frac{N_\beta^F}{Z_A(T)}
\pivec \exp\left[-\frac{1}{2}\intT A_\alpha (x)
A^\alpha (x)\right] Z_f[A;T,\mu]
\dle{auxpfmu}\\
Z_f[A;T,\mu]&=&\sumn \frac{m_0^{2n}}{(n!)^2}\pif 
\prodjint \sigma_+ (x_j) 
 \sigma_- (y_j)
\nonumber\\
&\times & \exp\left[\intT\bar\psi\left(\pabar+
ig\Abar (x)-\mu\gamma^0\right)\psi\right] .
\dle{zthmu1}
\ea
Notice that (\dr{auxpfmu}) and  (\dr{zthmu1}) are nothing but 
 the mass expansions of (\dr{mtgenfun2}) and (\dr{mtgenfun3}) respectively 
 for $\bar\eta=\eta=0$.  In order to calculate the  $m=0$
correlator in (\dr{zthmu1}), we put  back the external sources 
$\bar\eta$ and $\eta$ and take $m_0=0$  in (\dr{mtgenfun3}) and  
 recall the following result in  \dc{AlvGom98} 
\ba
Z_f^{m_0=0}[A,\bar\eta,\eta;T,\mu]&=&\frac{Z^F_0 (T,\mu)}{N_\beta^F}
\exp\left[-\intT\int_T d^2y \bar\eta (x)G(x,y,A;\mu)\eta(y)\right]\nonumber\\
&\times& 
\exp\left[-\frac{1}{2}\intT\int_T d^2y A_\alpha (x)\Pi^{\alpha\beta}(x-y)
 A_\beta (y)\right]
\nonumber\\
&\times &\exp\left[-i\frac{\mu g}{\pi}\intT A_0 (x)\right].
\dle{zfalvgom}
\ea
Here $Z^F_0 (T,\mu)$ is the $\mu\neq 0$ free massless fermion partition 
function which in 2D is given by
$$
\log Z^F_0 (T,\mu)=\log Z^F_0 (T,0)+\beta L\frac{\mu^2}{2\pi} \quad ,
$$
 and 
$$
\Pi_{\alpha\beta}(x-y)=\frac{g^2}{\pi}\fouri \left(\delta_{\alpha\beta}-
\frac{p_\alpha p_\beta}{p^2}\right)
$$
with $p_0=2\pi n T$, $n\in \IZ$.  $G(x,y,A,\mu)$ is the exact massless 
fermion propagator, which is given by \dc{AlvGom98}
\ba
G(x,y,A;\mu)&=&\exp\left\{ig[\chi (x)-\chi(y)]\right\}S(x,y;\mu)
\nonumber\\
\chi (x)&=&\int_T d^2 z \Delta_T (x-z) \pabar_z \Abar (z)
\nonumber
\ea
with $ \Delta_T$ the free boson propagator in (\dr{bosprop}) and 
$S(x,y;\mu)$ is the free fermion $\mu\neq 0$ propagator, solution of
$(\pabar-\mu\gamma^0)S(x,y;\mu)=\delta^{(2)} (x-y)$, for 
which we can write
$$
S(x,y;\mu)=\exp\left[-i(x^1-y^1)\mu\gamma^5\right] S(x,y;0).
$$
Now redefine fields and sources as $\bar\psi\rightarrow
\bar\psi\gamma^0$ and $\bar\eta\rightarrow \bar\eta\gamma^0$, so that
 $\bar\eta\psi\rightarrow \bar\eta_1\psi_2+\bar\eta_2\psi_1$ and 
$\bar\psi\eta\rightarrow \bar\psi_1\eta_2+\bar\psi_2\eta_1$ and, with
the above definitions, 
\ba
\bar\eta (x)G(x,y,A;\mu)\eta(y)&=&\sum_{a=1}^2
\bar\eta_a (x)G_a (x,y,A;\mu) \eta_a (y),
\nonumber\\
G_\pm (x,y,A;\mu)&=&-\frac{1}{2\beta}
\frac{Z_Q^\mp (x-y)}{Q^2 (x-y)}
\exp\left[\chi^\pm (x)-\chi^\pm (y)\mp i\mu (x^1-y^1)\right]\!\!,
\dle{zthmu2}
\ea
where $G_+\equiv G_1$,  $G_-\equiv G_2$,  
$Z_Q^\pm$ are given  in (\dr{zq}) and
$$
\chi^\pm (x)=ig\int_T d^2 z \Delta_T (x-z) \left[ H(z) \pm i E(z)\right] 
$$
with $H=\partial_0 A_0+\partial_1 A_1$ and 
$E=\partial_0 A_1-\partial_1 A_0$. 
 Now, let us decompose the vector
 field as in (\dr{vecfi}), but letting  now  $\Delta \phi (x)\neq 0$,  
(see our previous  discussion for the massless  case), so that we have
\ba
-\frac{1}{2}\intT\int_T d^2y A_\alpha (x)\Pi^{\alpha\beta}(x-y)
 A_\beta (y)&=&\frac{1}{2\pi}\intT \partial_\alpha\phi \partial^\alpha \phi
\nonumber\\
-\frac{1}{2}\intT A_\alpha (x)A^\alpha (x)&=&-\frac{1}{2g^2}\intT\left[
\partial_\alpha\eta \partial^\alpha\eta -
\partial_\alpha\phi \partial^\alpha\phi \right]
\nonumber\\
-i\frac{\mu g}{\pi}\intT A_0 (x)&=&-\frac{\mu}{\pi}\intT 
\frac{\partial}{\partial x^1}\phi (x)
\nonumber\\
g\int_T d^2 z \Delta_T (x-z) E(z)&=& -i\phi (x)
\dle{zthmu3}
\ea
and  $H$ will not play any r\^{o}le.

 The final step  in order to calculate the
$\sigma_+\sigma_-$ correlator in (\dr{zthmu1}), is to 
  differentiate (\dr{zfalvgom}) with respect to the sources
$\bar\eta_a$ and $\eta_a$, after redefining both fields and sources 
as indicated above, that is,
\ba
\ll T_c \prodj \sigma_+ (x_j)  \sigma_- (y_j)\gg^{m_0=0}_A \; \;
\rightarrow \; \;
\ll T_c \bar\psi_2 (x_j)\psi_1 (x_j)\bar\psi_1 (y_j)\psi_2 (y_j)
 \gg^{m_0=0}_A\nonumber\\
=\prodj \frac{\delta}{\delta \eta_1 (x_j)}\frac{\delta}{\delta 
\bar\eta_2 (x_j)}\frac{\delta}{\delta \eta_2 (y_j)}
\frac{\delta}{\delta \bar\eta_1 (y_j)}Z_f^{m_0=0}[A,\bar\eta,\eta;T,\mu]
\ea
where the subscript $A$ indicates the thermal expectation value in the 
 presence of $A_\mu$, according to (\dr{zthmu1}). Then, taking into account 
  (\dr{zthmu2}) and 
(\dr{zthmu3}), we  find
\ba
Z_{MT}(T,\mu)&=&J_A\frac{Z_0^F (T,\mu)Z_0^B (T)}{Z_A (T)}
\sumn \left(\frac{1}{n!}\right)^2
\left(\frac{m_0}{2\beta}\right)^{2n}
\nonumber\\
&\times& 
\pib 
\exp\left[\intT \left(\frac{1}{2\kappa^2} 
\partial_\alpha\phi (x)\partial^\alpha\phi (x) 
-\frac{\mu}{\pi}\frac{\partial}{\partial x^1}\phi (x)\right)\right]
\nonumber\\
&\times&  \prodjint 
\frac{\prod_{k<j}\left[
Q^2(x_j-x_k)Q^2(y_j-y_k)\right]}
{\prod_{k=1}^n\left[Q^2(x_j-y_k)\right]}
\nonumber\\
&\times&  
\exp\left\{2i\left[\phi (x_j)-\phi (y_j)-\mu\left(x^1_j-y^1_j\right)
\right]\right\}.
\dle{zthmu4}
\ea

At this point, and to solve the path integral in $\phi$ above, we 
make use of the  measure translation 
invariance argument discussed before and, accordingly, 
 we shift $\phi\rightarrow
i\kappa(\phi-i\mu\kappa x^1/\pi)$ so that the $\mu$ term is eliminated
in the boson action in (\dr{zthmu4}). Then, we  can  calculate the boson
correlator as in previous sections and finally obtain
\ba
Z_{MT}(T,\mu)&=&Z_{MT}^{m=0}(T,\mu)\sumn \left(\frac{1}{n!}\right)^2 
\left[\frac{m}{2\beta}\left(\frac{\rho}
{T}\right)^{\kappa^2/\pi}\right]^{2n}
\nonumber\\
&\times&\prodjint 
\frac{\prod_{k<j}\left[
Q^2(x_j-x_k)Q^2(y_j-y_k)\right]^{1-\kappa^2/\pi}}{
\prod_{k=1}^n\left[Q^2(x_j-y_k)\right]^{1-\kappa^2/\pi}}
\nonumber\\
&\times& 
\exp\left[2i\mu\left(1-\frac{\kappa^2}{\pi}\right)(x^1_j-y^1_j)\right]
\dle{zthmudef}
\ea
where  $Z_{MT}^{m=0} (T,\mu)$ is given by (\dr{zthalvgom}) (recall that 
$Z^B_0(T)=Z^F_0 (T)$). We readily check that the above yields 
(\dr{thparfun}) for $\mu=0$. Another consistency check is that the
\mbox{$\mu\neq 0$} 
partition function we obtain is real, which can be easily 
 checked in (\dr{zthmudef}) just by redefining integration variables 
 $x_j\leftrightarrow y_j$. Besides,  by taking only the lowest terms 
 in the $m$-expansion, we have numerically 
checked that it is also positive definite.
%\dnote{Did you say 
%there was a better
%way of doing this with the Coulomb gas?. R:You mean the in high $T$ limit, 
%as an external electric field?}

Now, we expand (\dr{zsgmu}) in $\alpha_0$ and perform the shift 
(\dr{mushift}), assuming the measure invariance. Then, using  
(\dr{zthalvgom}) and (\dr{equivconst1})-(\dr{equivconst2}),
we obtain exactly (\dr{zthmudef}) and then the equivalence
(\dr{parfuneqchempot})-(\dr{zsgmu}) follows.

% Also recall that the operators $(\dr{equivoper1})-(\dr{equivoper2})$ have 
% themselves zero charge, since, in the fermion case they have the same 
% fermion and anti-fermion content and for bosons they are invariant under 
% $\phi (x)\rightarrow \phi (x)+(2\pi n/\lambda)\theta(x)$ 
%with $n\in\IZ$, which  
% would increase $Q_K$ by one. 

\section{Conclusions and Outlook}

We have studied in detail bosonisation at finite temperature and nonzero
 fermion chemical potential 
 in the massive Thirring and sine-Gordon models. The main new 
results of this paper are the following: 
the extension of canonical methods to $T>0$ and
 $\mu=0$ to prove that the partition functions of the two models 
 are the same; the proof, via path integral methods of the equivalence
 between thermal averages of different sets of zero-charge 
operators in the two models; and the analysis 
 of the $\mu\neq 0$ case  where the main result  is a topological new 
 term in the 
 bosonised SG partition function, 
accounting for the existence of fermion modes in the 
 spectrum. All our results have been proved in perturbation theory around 
 the massless Thirring model or, equivalently, the free scalar theory.

 In the canonical operator 
 proof, the  use of the thermal normal ordering was crucial.  
Also our results for the operators (mass and current) 
 equivalences, as well as for the $T>0$, $\mu\neq 0$ partition function, 
 clearly show the existence of fermion excitations in the thermal bath. 
This constitutes the  thermal identification between sine-Gordon 
kinks and Thirring fermions, so generalising the existing knowledge about this 
 subject obtained when Statistical Mechanics features were not considered.  
In addition, the topological term obtained for $\mu\neq 0$ is the natural 
 analogue of a similar term  found in the effective low-energy action for 
 QCD at finite baryon density. 

Among the issues which we would like to investigate further in the
future is the semiclassical approach, which we have commented
 about in the text, especially in what concerns the $\mu\neq 0$ case and the 
 topological action. Also, as was noted in the introduction, we would
 like to exploit the  relationship between fermions and kinks, when these 
 are treated as topological defects, since we believe that 
these 2D systems could become  a very promising playground for testing 
 cosmological models. The experimental realisation on Josephson
 junctions could then provide precise laboratory tests  
 where, in particular, the results 
 obtained in this paper  could be physically probed. 
Other points of interest which 
which we are looking at in more detail at the moment  
 include a numerical analysis of  the free energy and the $\mu\neq 0$ 
 fermion density, 
 the high temperature limit of these systems,
%\dnote{and its analogy with the 
% one-dimensional Coulomb gas????} 
and their behaviour near 
 the point $\lambda^2=8\pi$.

\section*{Acknowledgments}

We thank Peter Landshoff both for useful arguments and for
originally drawing our attention in this direction.  We are grateful  to
Tim Evans and Ray Rivers for numerous helpful discussions and to 
  R.F.Alvarez-Estrada for useful suggestions. 
A.G.N.\ has received 
support through CICYT, Spain, project AEN97-1693 and through  
a fellowship of  MEC, Spain, and would like to thank the Imperial College 
Theory Group for their hospitality during the completion of this work.
D.A.S.\ is supported by P.P.A.R.C.\ of the UK through a
research fellowship and is a member of Girton College, Cambridge. 
This work was supported in part by the E.S.F.

%\newpage
\typeout{--- No new page for bibliography ---}

\end{document}